\documentclass[a4paper,11pt]{article}
\usepackage{jheppub} 
\usepackage{comment}

\title{\boldmath Giant Gravitons, Fermionic Forms and Vertex Algebras}

\author[a]{Nick Dorey,}
\author[a]{Paul Luis Roehl}
\affiliation[a]{Department of Applied Mathematics and Theoretical Physics,
University of Cambridge, Cambridge CB3 0WA, UK}

\emailAdd{nd217@cam.ac.uk}
\emailAdd{plr30@cam.ac.uk}

\abstract{We investigate the mathematical and physical content of the giant graviton expansion of three-dimensional $\mathcal{N}=4$ superconformal field theories in a simplifying limit. We uncover an interesting relation between the coefficients in this expansion, the Hilbert series of certain quiver varieties and the representation theory of vertex algebras.  In particular, for the worldvolume theory of $N$ M2-branes at the tip of a toric hyper-K\"{a}hler four-fold cone: $X_{4}=\mathbb{C}^2 /{\mathbb{Z}_L} \times \mathbb{C}^2/{\mathbb{Z}_K}$, we derive an explicit expression for the coefficients in terms of {\em affine fermionic forms} and show that they coincide with characters of a direct sum of parafermionic W-algebras.} 

\begin{document}
\maketitle
\flushbottom

\section{Introduction}
\paragraph{}
As holographic dualities typically relate the classical regime of 
bulk gravity  to a large-$N$/strongly-coupled limit of the boundary CFT, it is often hard to use them to extract information about the emergence of classical gravity and the form of quantum corrections. The superconformal index \cite{Kinney:2005ej, romelsberger2006, Bhattacharya:2008zy} provides a rare example of an observable in the boundary CFT which can be computed exactly and expanded to provide detailed information about the form of quantum corrections in the bulk. In particular, the Giant Graviton expansion \cite{Arai:2019xmp,Imamura:2021ytr,Gaiotto:2021xce, murthy2023} of the superconformal index admits an interpretation as a series of corrections arising from the contribution of flux-stabilized branes in the dual geometry \cite{McGreevy:2000cw, grisaru2000, hashimoto2000}.  
In the following, we will focus on a family of $\mathcal{N}=4$ supersymmetric  gauge theories in three dimensions which are realised on the worldvolume of $N$ M2-branes at the tip of a hyper-K\"{a}hler fourfold cone: $X_{4}=\mathbb{C}^2 /{\mathbb{Z}_L} \times \mathbb{C}^2/{\mathbb{Z}_K}$ \cite{benini2010}. These theories are holographically dual to M-theory on $\text{AdS}_4\times \text{SE}_{7}$, where $\text{SE}_{7}$ is the Tri-Sasakian manifold corresponding to the base of the cone $X_4$. In this context, giant graviton corrections correspond to the contribution of M5-branes wrapped on topologically trivial cycles\footnote{There are also contributions from M5-branes wrapped on topologically stable cycles corresponding to baryons in the dual field theory.} in $\text{SE}_{7}$. The form of the expansion suggests interesting new relations between the observables of the original boundary CFT on the M2-branes and those of the world-volume theory on wrapped M5-branes which we will investigate further in this paper.
\paragraph{}
The full superconformal index, although known in closed form, is of limited use for the purposes of this paper. In particular, it has not so far been possible to derive the giant graviton expansion directly by expanding the full index. For this reason, we will focus throughout on simplifying limits \cite{Razamat:2014pta} in which the superconformal index reduces to the Hilbert series which counts holomorphic functions\footnote{In this limit, baryon numbers are realised as fluxes for certain global symmetries which can also be incorporated by modifying the Hilbert series to count holomorphic sections of an appropriate bundle rather than functions.} on the Higgs branch of the vacuum moduli space\footnote{One can also consider a corresponding limit in which the superconformal index reduces to the Hilbert series of the Coulomb branch. As the $\mathcal{N}=4$ theories considered in this paper exhibit three-dimensional mirror symmetry \cite{intriligator1996}, the Higgs and Coulomb branch limits of mirror pairs coincide.}. With this simplification, we will obtain an explicit expression for the coefficients in the giant graviton expansion. Strikingly, they are given by the affine versions of the {\em fermionic forms}
which appear pervasively in studies of the representation theory of quantum groups and the geometry of quiver varieties \cite{Hatayama:2001gy}. In some cases, we can reproduce these coefficients directly from the M5 worldvolume theory, but more generally we can relate them to characters of an auxiliary vertex algebra. In particular, the giant graviton coefficients of the most general case described above all correspond to particular characters of a certain direct sum of parafermionic W-algebras.
The appearance of W-algebra characters here seems to be closely connected to the AGT Correspondence \cite{alday2010} and other related manifestations of vertex algebras in supersymmetric gauge theory and string theory \cite{Gaiotto:2017euk, Beem:2013sza, Beem:2014kka}.
In the remainder of this introductory section, we describe our main results in more detail.

\paragraph{}
For $L = K = 1$, the world volume theory of $N$ M2-branes at the origin of $X_4 = \mathbb{C}^4$ coincides with the ABJM superconformal field theory \cite{Aharony:2008ug} in $D=3$ (at level $k=1$) \cite{Bashkirov:2010kz}, which is dual to M-theory on $\text{AdS}_4 \times S^7$. 
The superconformal index \cite{Kim:2009wb, Imamura:2011su} of this theory admits a giant graviton expansion of the form, \cite{Arai:2020uwd}
\begin{equation}
  \label{eq:2510121528}
	\mathcal{I}_N(y, q_I) = \mathcal{I}_{\infty}(y, q_I)
	\sum_{n_I = 0}^{\infty} \left(\prod_{I=1}^{4} q_I^{N n_I}\right) Z_{n_1, n_2, n_3, n_4}(y, q_I).
\end{equation}
Here, $q_{I}$ for $I=1,2,3,4$ are fugacities for the four commuting $U(1)$ R-symmetries of the ABJM theory corresponding to the isometries of $\mathbb{C}^{4}$, while $y$ is a fugacity for angular momentum.
The giant graviton configurations on $S^7$ are labelled by four integers. Their corresponding coefficients
$Z_{n_1, n_2, n_3, n_4}$ are interpreted as the contribution of $1/8$-BPS configurations \cite{mikhailov2000} in which M5-branes are multiply-wrapped on a basis set of four contractible $S^5$ cycles\footnote{In more detail, the resulting cycle corresponds to the intersection of the holomorphic divisor $\{z_1^{n_1} z_2^{n_2} z_3^{n_3} z_4^{n_4} = 0\}$ of $\mathbb{C}^{4}$ with $S^7 = \{|z_1|^2 + |z_2|^2 + |z_3|^2 + |z_4|^2 = 1\}$. In this picture, a number $n_I > 0$ corresponds to $n_I$ M5-branes wrapped on the hypersurface $z_I = 0$ of $S^7$.} in $S^{7}$.
Imamura et al. have proposed that the corresponding coefficient function, $Z_{n_1, n_2, n_3, n_4}$, can itself be interpreted as a suitable index for the worldvolume theory of the wrapped M5-branes \cite{Imamura:2021ytr, Arai:2020uwd}.  
\paragraph{}
In fact, it has been observed \cite{Gaiotto:2021xce, Imamura:2022aua} that, because of cancellations between positive and negative terms, the giant graviton expansion has many inequivalent forms. These cancellations reflect an underlying gauge invariance in the problem related to the presence at finite $N$ of the trace constraints in the dual field theory. 
Remarkably, this phenomenon effectively reduces \eqref{eq:2510121528} from four summation variables
to a single one, giving
\begin{equation}
  \label{eq:2510121535}
	\mathcal{I}_N(y, q_I) = \mathcal{I}_{\infty}(y, q_I)
	\sum_{n_1 = 0}^{\infty} q_1^{N n_1} Z_{n_1, 0, 0, 0}(y, q_I).
\end{equation}
This reduction is referred to as a simple-sum expansion. According to Imamura's proposal, the remaining coefficients $Z_{n_1, 0, 0, 0}$ should correspond (after analytic continuation) to the superconformal indices of the $A_{n}$ series of 
$(2,0)$ SCFTs in six dimensions (with $n=n_{1}-1$). We will make contact with this proposal below. 

\paragraph{}
The story described above has a natural generalisation \cite{Arai:2019aou} to a much larger class of three-dimensional field theories, which arise when the $\mathbb{C}^4$ space transverse to the M2 branes is replaced by a general toric Calabi-Yau fourfold cone.
The Calabi-Yau condition together with conical structure guarantees that the resulting field theory has $\mathcal{N}=2$ superconformal invariance in three dimensions.
The toric condition ensures that, as for the ABJM case, the theory has four commuting $U(1)$ symmetries corresponding to the isometries of $X_{4}$.
The $U(1)_{R}$ symmetry of the $\mathcal{N}=2$ superconformal algebra is realised as a particular linear combination of these ``mesonic'' symmetries.
In the holographic dual the compact internal space $S^{7}$ is replaced by a non-trivial Sasaki-Einstein $7$-manifold $SE_{7}$, which typically has non-contractible cycles.
These give rise to additional ``baryonic'' symmetries in the dual field theory.
Field theory states carrying the corresponding baryonic charges correspond to M5-branes wrapped on non-contractible five-cycles in $SE_{7}$.
In these theories giant graviton and baryonic configurations of the M5 brane can be understood in parallel as supersymmetric cycles\footnote{If we realize $X_{4}$ as a cone over $SE_{7}$, then the recipe is the same as that described in the $\mathbb{C}^{4}$ case above: a supersymmetric cycle corresponds to the intersection $D\cap SE_{7}$ where $D$ holomorphic divisor of $X_{4}$.} in $SE_{7}$.         
\paragraph{}
In this paper we focus on the slightly less-general case where $X_{4}$ is hyper-K\"{a}hler and $\mathcal{N}=4$ supersymmetry is preserved in three dimensions. 
The most general toric hyper-K\"{a}hler cone of dimension four is a product of two ALE singularities. Thus, we consider the $\mathcal{N}=4$ superconformal field theory arising on the world volume of $N$ M2-branes at the origin of $X_{4}=\mathbb{C}^2 / \mathbb{Z}_L  \times \mathbb{C}^2 / Z_K$ singularity. We will denote this theory as $\mathcal{T}_{N}[K,L]$. Counting the inequivalent holomorphic divisors of $X_{4}$, we deduce that this theory has $L+K-2$ baryonic symmetries in addition to the four mesonic ones corresponding to the isometries of $X_{4}$.

\begin{figure}[]
	\centering
  \includegraphics{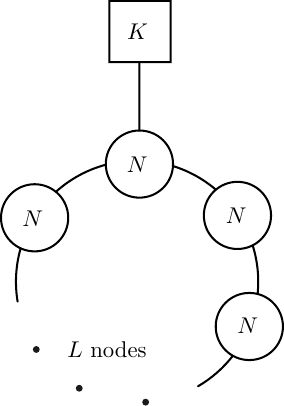}
  \caption{The quiver diagram for the gauge theory $\mathcal{T}_N[K,L]$.
    Circular nodes correspond to $U(N)$ gauge groups. The square node is a $U(K)$ flavour group.
		Lines between nodes correspond to bi-fundamental hypermultiplets.\label{fig:quiver_diagram}}
\end{figure}

\paragraph{}To study $\mathcal{T}_{N}[K,L]$ it is convenient to consider a weakly-coupled three-dimensional gauge theory which flows to this superconformal fixed point in the IR. As the supersymmetric indices remain constant along RG flows, the superconformal index of $\mathcal{T}_{N}[K,L]$ can then be calculated directly in the 3D gauge theory description which is weakly coupled in the UV. The full RG flow can be realised directly in the M-theory construction by replacing the ALE factor $\mathbb{C}^2 / \mathbb{Z}_K$ factor in the spacetime by a corresponding ALF geometry and applying Type IIA/M-theory duality to the resulting background.\footnote{Note that this deformation breaks the manifest symmetry of $\mathcal{T}_{N}[K,L]$ under the interchange of $K$ and $L$. In the dual gauge theory, the recovery of this invariance in the IR is the essential content of three-dimensional mirror symmetry.}
The corresponding IIA solution includes $N$ D2 branes and $K$ D6 branes, the latter also filling the remaining singular factor $\mathbb{C}^2 / \mathbb{Z}_L$. By standard arguments the worldvolume of the brane intersection is described by an affine quiver gauge theory. The gauge group includes $U(N)$ factors arranged in a circle corresponding to $L$ nodes of the quiver diagram in Figure \ref{fig:quiver_diagram}. In the following, we introduce a $\mathbb{Z}_{L}$ index, $A=0,1,\ldots,L-1$ (identified modulo $L$) labelling the nodes or gauge-group factors. The matter content includes bifundamental hypermultiplets corresponding to lines in the diagram joining adjacent nodes. There are also a total of $K$ additional hypermultiplets, each in the fundamental representation of (one of) the $U(N)$ gauge-group factors. We will also introduce a $\mathbb{Z}_{K}$ index $\alpha=0,1,\ldots,K-1$ (identified modulo $K$) labeling the flavours

 \paragraph{}
 The gauge theory has a vacuum moduli space which includes Higgs, Coulomb and various mixed branches. The Higgs branch is realised straightforwardly as a hyper-K\"{a}hler quotient. The resulting complex space determined by the data described above, is known as a Nakajima quiver variety. In the present case, the Higgs branch also corresponds, via the ADHM construction, to the moduli space $\mathcal{M}_{N}[K,L]$ of $N$ instantons in a $U(K)$ Yang-Mills theory defined on $\mathbb{C}^{2}/\mathbb{Z}_{L}$. In the special case of (non-commutative) abelian instantons, $K=1$, the moduli space is (a resolution of) the symmetric product of $N$ copies of  $\mathbb{C}^{2}/\mathbb{Z}_{L}$. The Coulomb branch of $\mathcal{T}_{N}[K,L]$ is more subtle due to quantum corrections, but can be studied using three-dimensional mirror symmetry as we describe below.    
The $L+K+2$ global symmetries of $\mathcal{T}_{N}[K,L]$ are also present in the gauge theory description.
They naturally split into $K+1$ conventional global symmetries carried by the elementary quanta of the weakly-coupled gauge theory, and a further $L+1$ topological symmetries which are carried by gauge theory vortices.
On the vacuum moduli space, the first set are realised geometrically as isometries of the Higgs branch, while the second correspond to isometries of the Coulomb branch.
Further symmetry enhancement occurs at the singular point where the branches intersect: $K-1$ commuting $U(1)$ factors are enlarged to an $SU(K)$ global symmetry.
Similarly, mirror symmetry implies that $L-1$ of the topological $U(1)$ symmetries are enhanced to an $SU(L)$ invariance.
Of the original four ``mesonic'' symmetries of $\mathcal{T}_{N}[K,L]$, the two with fugacities $q_{1}$ and $q_{2}$ are Higgs branch symmetries in this sense, those with fugacities $q_{3}$ and $q_{4}$ are topological symmetries.
We will introduce a fugacity $x_{\alpha}$, $\alpha=0,1,\ldots,K-1$ for the global symmetry associated with each flavour and $y_{A}$, $A=0,1,\ldots,L-1$ for the topological symmetry associated with each gauge group factor. 
We denote them collectively by $\vec{x}$ and $\vec{y}$.
The fugacities are subject to the constraints
\begin{equation}
  \label{eq:2510311953}
(q_1 / q_2)^{L/2} = \prod_{\alpha = 0}^{K-1} x_{\alpha} \equiv x
 \quad \text{and} \quad 
 (q_3 / q_4)^{K/2} = \prod_{A=0}^{L-1} y_A \equiv y.
\end{equation}
 \paragraph{}
 The superconformal index counts local operators or states on $S^{2}$, graded by their charges under each of the symmetries described above.  There are also additional sectors corresponding to the introduction of magnetic fluxes for each global symmetry. For the topological charges introduced above, the corresponding fluxes can also be interpreted as baryon number corresponding to sectors of non-zero charge under the $U(1)$ center of each of the $L$ $U(N)$ gauge group factors\footnote{From now on the term baryon number will refer explicitly to charge under these symmetries in contrast with the more general notion of baryonic symmetries used in the introductory paragraphs above.}.   
 \paragraph{}
 In this paper, we will consider the giant graviton expansion of the superconformal index in sectors of fixed global magnetic flux/baryon number.  
 As explained above, it is hard to analyse the index in full generality. However, we can simplify the problem by taking a Higgs branch limit \cite{Razamat:2014pta} . The limit is achieved by taking $y, q_{3}, q_{4}\rightarrow 0$, which kills the contribution of any states carrying the corresponding charges, with $q_{1}$ and $q_{2}$ fixed.  
 In the absence of fluxes, the superconformal index reduces in this limit  to the Hilbert series which counts holomorphic functions on the complex space $\mathcal{M}_{N}[K,L]$. Additionally, one can introduce baryonic charge $\mathbf{B}$ for the central $U(1)^L \subset U(N)^L$ into the Hilbert series by modifying it to count holomorphic sections of certain bundles defined over $\mathcal{M}_{N}[K,L]$. As only the Higgs branch symmetries identified above act non-trivially, the limiting of the index depends only on the $x_{\alpha}$ (in addition to $q_{1}$ and $q_{2}$). It has no dependence on the Coulomb branch parameters $y_{A}$. Although states electrically charged under the corresponding symmetries decouple in this limit, the corresponding magnetic fluxes are unsuppressed and can be identified with baryon number as described above. 
 \paragraph{}
 In the symmetric product case $K=1$, the Higgs branch Hilbert series can easily be extracted via its generating function which has a simple plethystic form. As we review below, it is then straightforward to determine the corresponding giant graviton coefficients. For $K>1$, the problem is harder but we can make progress by using the fact that $\mathcal{M}_{N}[K,L]$ can also be identified as the Coulomb branch of the mirror theory $\mathcal{T}_{N}[L,K]$ and its Hilbert series can be evaluated using the monopole formula \cite{Cremonesi:2013lqa, cremonesi2014a, Cremonesi:2014xha} of Hanany et al. Importantly, we discover that the monopole formula undergoes significant simplifications in the large $N$ limit which allow us to extract the giant graviton coefficients in closed form. Further, as 3D mirror symmetry interchanges conventional global symmetries with topological ones, one can also incorporate non-zero baryon numbers by introducing fluxes for the global symmetries of the mirror theory.   
\paragraph{}
 
    
\begin{figure}[]
	\centering
  \includegraphics{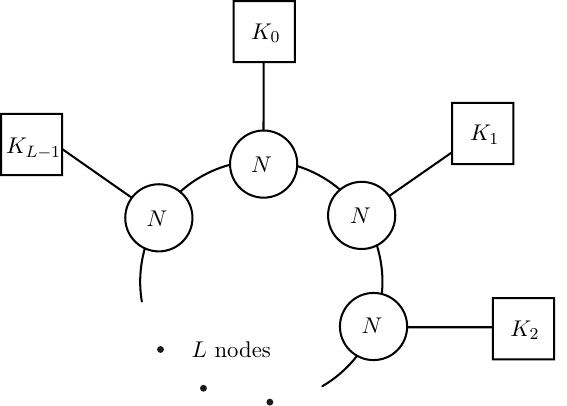}
  \caption{The quiver diagram for the class of gauge theories generalising $\mathcal{T}_N[K,L]$.
    There are $K$ total flavours, distributed onto the $L$ nodes as $\sum_{A = 0}^{L-1} K_A = K$.
		\label{fig:quiver_diagram_general_flavours}}
\end{figure}
To define the index in the presence of both global and topological fluxes, it is convenient to work with a generalisation of the quiver gauge theory described above where the $K$ fundamental multiplets are shared between the $L$ nodes of the quiver diagram, see Figure \ref{fig:quiver_diagram_general_flavours}. We will consider the case with $K_{A}$ hypermultiplets in the fundamental representation of the $U(N)$ gauge group factor associated with the $A$'th node for $A=0,1,\ldots L-1$ where $\sum_{A = 0}^{L-1} K_{A} = K$. The Higgs branch of this theory is equivalent to the Coulomb branch of a mirror quiver is a $U(N)^K$ circular quiver gauge theory with $L$ total flavours distributed between the $K$ nodes. Specifically there are $L_{\alpha}$ flavours at 
the $\alpha$'th node for $\alpha=0,1,\ldots,K-1$ with $\sum_{\alpha = 0}^{K-1} L_{\alpha} = L$. The mirror map completely determines the $\{L_{\alpha}\}$ in terms of the 
$\{K_{A}\}$ \cite{deBoer:1996mp}.\footnote{First, rotate the quiver such that $K_0 \not= 0$. Then, consider a Young diagram $Y$ of length $L$. The rows are given by $Y_{i} = \sum_{A = 0}^{L-i} K_A$, $i = 1, \ldots, L$. The transpose diagram $Y^{T}$ has length $K$ and the $L_{\alpha}$ are such that $Y_{i}^T = \sum_{\alpha = 0}^{K-i} L_{\alpha}$, $i = 1, \ldots, K$.}
Individual flavours are labelled by a double index $(\alpha,a)$ with $\alpha = 0, \ldots, K-1$, $a = 1, \ldots, L_{\alpha}$. The most general observable we will study below is the Hilbert series of the Coulomb branch of this mirror theory in the presence of $\mathbf{M}_{\alpha, a}$ units of flux for the corresponding global symmetry. In the orginal theory, this corresponds to the Higgs branch limit of the superconformal index with generic global fluxes and baryon numbers turned on. We will comment on the global fluxes of the original theory in Section \ref{sec:general_quiver}. The baryon numbers are encoded in the $L$ component vector $\mathbf{M} = (\mathbf{M}_{\alpha, a}) \in \mathbb{Z}^L$.
Using the action of the Weyl group of the unbroken symmetry, we can permute the fluxes to place them in decreasing order at each node, choosing $\mathbf{M}_{\alpha, a}$ for each fixed $\alpha$ such that $\mathbf{M}_{\alpha,1} \geq \mathbf{M}_{\alpha,2} \geq \ldots \geq \mathbf{M}_{\alpha, L_{\alpha}}$.
Furthermore, the central $U(1)$ subgroup of the baryonic symmetry decouples, leaving the Hilbert series invariant under a simultaneous shift $\mathbf{M}_{\alpha, a} \to \mathbf{M}_{\alpha, a} + m$ for $m \in \mathbb{Z}$. 
We denote by $[\mathbf{M}]$ the equivalence class of such ordered vectors $\mathbf{M}\in \mathbb{Z}^{L}$ under the shift action. The resulting Hilbert series is then denoted
$H_N^{[\mathbf{M}]}(q_1, q_2; \vec{x})$. The starting point of our investigation is the explicit expression for this quantity provided by the monopole formula which is given in eq.~\eqref{eq:2509251529} below.
\paragraph{}
After this lengthy set-up, we can now describe our main results. We argue on general grounds that the Hilbert series has a single sum giant graviton expansion of the form, 
\begin{equation}
  H_N^{[\mathbf{M}]}(q_1, q_2; \vec{x}) =
H_{\infty}(q_1, q_2; \vec{x}) \sum_{\mathbf{m} \in [\mathbf{M}]_+} q_2^{N |\mathbf{m}|}
  Z_{\mathbf{0},\mathbf{m}}(q_1, q_2; \vec{x})
\end{equation}
when expanded around $q_1 = 0$ and 
\begin{equation}
H_N^{[\mathbf{M}]}(q_1, q_2; \vec{x}) =
  H_{\infty}(q_1, q_2; \vec{x}) \sum_{\mathbf{m} \in [-\mathbf{M}]_+} q_1^{N |\mathbf{m}|}
  Z_{\mathbf{m},\mathbf{0}}(q_1, q_2; \vec{x})
\end{equation}
when expanded around $q_2 = 0$. Here,
\begin{equation}
|\mathbf{m}| \equiv \sum_{\alpha = 0}^{K-1} \sum_{a=1}^{L_{\alpha}} \mathbf{m}_{\alpha, a}
\end{equation}
and $[\mathbf{\pm M}]_+$ denotes those elements in the equivalence class $[\pm \mathbf{M}]$ with purely non-negative entries, i.e.~$\mathbf{m}_{\alpha, a} \geq 0$ for $\mathbf{m} \in [\pm \mathbf{M}]_+$. For $[-\mathbf{M}]$, the entries are once again assumed to be sorted in descending order, using the action of the Weyl group.
\paragraph{}
Assuming the above form of the giant graviton expansion, our main result is the proof that the coefficients are given by
\begin{align}
  \label{eq:2509290955}
  Z_{\mathbf{0},\mathbf{m}}(q_1, q_2; \vec{x}) &=
(q_1^L; q_1^L)_{\infty} \;
  \frac{n(\mathbf{m}^T, q, \vec{q}_1^L)}{\Delta_{\widehat{\mathfrak{su}}(K)}(\vec{q}_1^L)} \,
     \quad \text{for} \quad \mathbf{m} \in [\mathbf{M}]_+ \quad \text{and} \\
  \label{eq:2510311206}
  Z_{\mathbf{m},\mathbf{0}}(q_1, q_2; \vec{x}) &=
(q_2^L; q_2^L)_{\infty} \;
  \frac{n(\mathbf{m}^T, q, \vec{q}_2^L)}{\Delta_{\widehat{\mathfrak{su}}(K)}(\vec{q}_2^L)} \,
    \quad \text{for} \quad \mathbf{m} \in [-\mathbf{M}]_+.
\end{align}
Here,
\begin{equation}
    (a; y)_{\infty} \equiv \prod_{k=0}^{\infty} (1 - a y^k)
\end{equation}
is the Pochhammer symbol.
In order to define the other objects appearing in the above formula, we need to introduce some auxiliary concepts. A {\em quiver diagram} is a connected unoriented graph. Each such quiver diagram can be used to define a {\em generalised Kac-Moody algebra}.
In the simplest case, this is just the well known correspondence between simple Lie algebras and their Dynkin diagrams.
In the general case, the adjacency matrix of the quiver graph can be used to define a generalised Cartan matrix in a similar way.
With some additional data known as {\em framing}, each quiver diagram also defines a family of complex spaces known as {\em quiver varieties} \cite{nakajima1994, nakajima2002}.
The same data also defines a family of quiver gauge theories like the ones discussed above and the Higgs branch of the gauge theory coincides with the quiver variety.  
\paragraph{}
The objects appearing in (\ref{eq:2509290955}) can be defined in terms of the data described above for any quiver. In particular $\Delta$ is essentially the {\em Weyl denominator} of the corresponding generalised Kac-Moody algebra $\mathcal{A}$,
\begin{equation}
\Delta_{\mathcal{A}}(\vec{y}) =
\prod_{\alpha > 0}^{} (1 - y^{\alpha}),
\end{equation}
where the product is over all positive roots $\alpha$ of $\mathcal{A}$.
For our purposes, the relevant case is that of circular quivers with $K$ nodes for which the associated algebra is the affine Lie algebra $\hat{A}_{K-1}=\widehat{\mathfrak{su}}(K)$. As usual the nodes of the quiver correspond to simple roots of the algebra. 
Let $y_{\alpha}$, $\alpha = 0, \ldots, K-1$, be the fugacities for the simple roots.
For $Z_{\mathbf{0}, \mathbf{m}}$ and $Z_{\mathbf{m}, \mathbf{0}}$, these are the vectors $\vec{q}_1^L$ and $\vec{q}_2^L$ respectively, whose components are defined as
\begin{equation}
  q_{1, \alpha}^L \equiv q^{L_{\alpha} / 2} x_{\alpha}
  \quad \text{and} \quad
  q_{2, \alpha}^L \equiv q^{L_{\alpha} / 2} x_{\alpha}^{-1},
  \quad \text{where} \quad
  q = q_1 q_2.
\end{equation}
The Weyl denominator $\Delta_{\widehat{\mathfrak{su}}(K)}(\vec{y})$ is best written down in terms of the fugacity $y$ for the imaginary root and fugacities $z_{\alpha}$, $\alpha = 1, \ldots, K$ for the non-affine $\mathfrak{su}(K)$, subject to the constraint $\prod_{\alpha = 1}^{K} z_{\alpha} = 1$.
These are related to $y_{\alpha}$ by
\begin{equation}
  y = \prod_{\alpha = 0}^{K-1} y_{\alpha}
  \quad \text{and} \quad
  y_{\alpha} = z_{\alpha} / z_{\alpha+1}
  \quad \text{for} \quad
  \alpha = 1, \ldots, K-1.
\end{equation}
In particular, $y = q_1^L$ or $y = q_2^L$ for \eqref{eq:2509290955} and \eqref{eq:2510311206}.
The Weyl denominator is then
\begin{equation}
    \Delta_{\widehat{\mathfrak{su}}(K)}(\vec{y}) =
    (y; y)_{\infty}^{K-1}
    \prod_{\alpha < \beta}^{K}
    (z_{\alpha}/z_{\beta}; y)_{\infty}
    \prod_{\alpha > \beta}^{K}
    (y z_{\alpha}/z_{\beta}; y)_{\infty}.
\end{equation}
\paragraph{}
Similarly, the function $n(\nu, q, \vec{y})$ is defined for an arbitrary generalised Kac-Moody algebra as 
\begin{equation}
  \label{eq:2510141337}
	n(\nu, q, \vec{y}) \equiv \sum\limits_{\tau \in \mathcal{P}^K}
  \prod_{k = 1}^{\infty} q^{-(\nu_k, \tau_k)} q^{\frac{1}{2}(\tau_k, \tau_k)}
  \prod_{\alpha = 0}^{K-1}
  y_{\alpha}^{\tau_k^{\alpha}}
	\left[\sum\limits_{a=1}^k (\nu_{\alpha, a} - \tau_{\alpha, a}), \tau_k^{\alpha} - \tau_{k+1}^{\alpha} \right]_q.
\end{equation}
Specialising to the case of the affine Lie algebra $\widehat{\mathfrak{su}}(K)$,
the transposed background charge $\nu$ is interpreted as a list of non-negative \emph{weight vectors} of
$\widehat{\mathfrak{su}}(K)$, $\nu_k = (\nu_{\alpha, k})_{\alpha = 0, \ldots, K-1} \in \mathbb{N}_0^K$.
The summation is over $K$-tuples of partitions $\tau$, which are identified with non-negative \emph{root vectors} $\tau_k = (\tau_k^{\alpha})_{\alpha = 0, \ldots, K-1} \in \mathbb{N}_0^K$.
The inner product between a root and a weight (or two roots) is
\begin{equation}
  (\nu_k, \tau_k) = \nu_{\alpha, k} \tau^{\alpha}_k
  \quad \text{and} \quad
  (\tau_k, \tau_k) = C_{\alpha \beta} \tau_k^{\alpha} \tau_k^{\beta}
\end{equation}
with an implicit summation over $\alpha$ and $\beta$.
The matrix $C_{\alpha \beta} = 2 \delta_{\alpha \beta} - \delta_{\alpha, \beta + 1} - \delta_{\alpha, \beta - 1}$ is the $K \times K$ generalised Cartan matrix of $\widehat{\mathfrak{su}}(K)$.
Here, the indices on the Kronecker delta functions are understood as modulo $K$.
A root vector can be converted into a weight vector by multiplication with the generalised Cartan matrix,
\begin{equation}
  \tau_{\alpha, k} = C_{\alpha \beta} \tau_k^{\beta}.
\end{equation}
Since $C$ is degenerate for affine Lie algebras, this procedure is not invertible.
Finally, the quantity
\begin{equation}
  [n, m]_q = {n+m \choose m}_q = \prod_{i=1}^{m} \frac{(1-q^{n+i})}{1-q^{i}}
  \quad \text{for} \quad
  n \in \mathbb{Z}, m \in \mathbb{N}_0.
\end{equation}
is called the $q$-binomial coefficient.

\paragraph{}
The combination which makes up the right-hand sides of equations (\ref{eq:2509290955}, \ref{eq:2510311206}) are both examples of so-called {\em fermionic forms} \cite{Hatayama:2001gy, mozgovoy2007ffqv}. In the special case where $\mathcal{A}$ is a simple Lie algebra, these originated in the study of integrable spin chains. In this context, they are also closely related to $q$-characters of quantum algebras and this leads to their most striking property: their Taylor series expansions in powers of $q$ and $y_{\alpha}$ contain only non-negative integer coefficients. As mentioned, this highly non-manifest positivity has a representation theoretic origin: the resulting expressions represent the $q$-dimensions of weight subspaces in a module of the corresponding Yangian $\mathbb{Y}[\mathcal{A}]$. This connection is by now well established when $\mathcal{A}$ is a simple Lie algebra. For example, the exact analogue of \eqref{eq:2509290955} for a linear quiver corresponding to an $\mathfrak{su}(K)$ Dynkin diagram is the character of the fusion product of certain classical Kirillov-Reshetikhin modules \cite{di2008proof}. Our results address the much less explored case where $\mathcal{A}$ is an affine Lie algebra. Once again we can demonstrate positivity of the coefficients. This strongly suggests an interpretation in terms of the corresponding modules of the {\em affine Yangian} $\mathbb{Y}[\widehat{\mathfrak{su}}(K)]$. We leave this for future investigation but, in the following, provide a more limited interpretation which holds in the classical limit $q\rightarrow 1$. In particular, in this case the affine Yangian is known to degenerate to an ordinary vertex algebra $\mathcal{V}$ of $W$-type and we can recover the corresponding characters explicitly.   
\paragraph{}
Positivity in the sense described above can also be understood more directly by showing that the fermionic form provides a $q$-counting of something. 
We claim that the giant graviton coefficient $Z_{\mathbf{0}, \mathbf{m}}$ $q$-counts $L$-tuples of coloured plane partitions with height restrictions given by $\mathbf{m}$\footnote{More precisely, we mean that $Z_{\mathbf{0}, \mathbf{m}}$, when expanded in $\vec{q}_{1}$, naturally arranges itself as an expansion in the fugacities $q_{2,\alpha}^{-1} = q^{-L_{\alpha}} q_{1, \alpha}$ whose coefficients are polynomials in $q$ with positive integer coefficients. Every monomial $q^{\#}$ corresponds to exactly one configuration of plane partitions. For $\mathbf{m}_{\alpha,a} \leq 1$, this can be seen explicitly in eq.~\eqref{eq:2511021507}.} (see also \cite{Dijkgraaf:2007fe}).
Concretely, for given background charge $\mathbf{m}$, every value $\mathbf{m}_{\alpha, a}$
corresponds to a plane partition $Y^{(\alpha)}_{\mathbf{m}_{\alpha,a}}$ of starting colour $\alpha \in \mathbb{Z}_K$ and maximal height $\mathbf{m}_{\alpha,a}$.
The colouring associates to any box $(i,j,k) \in Y_{\mathbf{m}_{\alpha,a}}^{(\alpha)}$
(with $k \leq \mathbf{m}_{\alpha,a}$) the colour $\alpha + i - j \mod K$.
\paragraph{}
The fermionic form determined by quiver data also has an interpretation in terms of the geometry of the corresponding quiver variety. More precisely, in the special case where each of the plane partitions is of maximal height one and becomes an ordinary partition, the quantity $Z_{\mathbf{0}, \mathbf{m}}$ is precisely the generating function for the Poincar\'e polynomials of this family of quiver varieties \cite{Hausel_2006, mozgovoy2007ffqv}. This connection is very suggestive of a direct explanation of our results in terms of the worldvolume dynamics of the M5 brane. 
\paragraph{}
We first describe these various connections in the simplest case $L = K = 1$ corresponding to the ABJM SCFT. In this case we show in Appendix \ref{sec:fermionic_forms-adhm_quiver}
how equation \eqref{eq:2509290955} recovers the known form of the giant graviton coefficients \cite{Gaiotto:2021xce, Hayashi:2024aaf}
\begin{equation}
  Z_{0,m}(q_1, q_2) = \prod_{l=1}^{m} \frac{1}{(q_2^{-l}; q_1)_{\infty}}.
\end{equation}
This result can be understood directly from the M5 world volume theory, which is the $(2,0)$ superconformal theory of type $A_{m-1}$.
More precisely, Imamura's proposal predicts that the same expression should be recovered from a specific limit of the $(2,0)$ superconformal index. Following \cite{Hayashi:2024aaf} we will explicitly confirm this expression in Section \ref{sec:twisted_limit} below.    
In the unrefined limit $q = q_1 q_2 \to 1$, the coefficients further reduce to the vacuum character of the $W$-algebra
$\mathcal{V} = \mathcal{H} \oplus W_m$ with central charge $c(W_m) = m-1$ and $c(\mathcal{V}) = m$. This connection is well established on the M5 brane side where the algebra in question coincides with the chiral algebra of the $(2,0)$ theory \cite{Beem:2014kka, Bullimore:2014upa} and the sphere partition function is known to reduce to the vacuum character of $\mathcal{V}$.     
\paragraph{}
Focussing on the theory $\mathcal{T}_{N}[K,1]$ with $K > 1$, the coefficients \eqref{eq:2509290955} depend on a
single giant graviton number $m \in \mathbb{N}_0$.
When $q \to 1$, we have checked to high orders that the resulting coefficient has the combinatorial interpretation of counting $K$-coloured plane partitions of maximal height $m$. Remarkably, Manabe \cite{Manabe:2020etw} discovered that same generating function 
coincides with a 
character of a certain coset algebra; 
\begin{equation}
  \label{eq:full_algebra}
	\mathcal{V}(m,K; p) = \mathcal{H} \oplus \widehat{\mathfrak{su}}(K)_m
	\oplus \underbrace{\frac{\widehat{\mathfrak{su}}(m)_K \oplus \widehat{\mathfrak{su}}(m)_{p-m}}
		{\widehat{\mathfrak{su}}(m)_{K+p-m}}}_{\equiv W_{m,K}^{\text{para}}}.
\end{equation}
Here $W_{m,K}^{\text{para}}$ is the $K$-th parafermion $W_m$-algebra.
When $K = 1$, this reduces back to $\mathcal{H} \oplus W_m$.
For $p \in \mathbb{N}$, the coset algebra describes $(p, p + K)$-minimal models of the parafermion $W$-algebra
$W_{m,K}^{\text{para}}$ of central charge
\begin{equation}
  c(W_{m,K}^{\text{para}}) = \frac{K (m^2 - 1)}{m + K} \left(1 - \frac{m(m+K)}{p(p+K)}\right).
\end{equation}
The characters of $\mathcal{V}(m, K; p)$ have the combinatorial interpretation as sums over $m$-tuples of partitions,
subject to a set of restrictions called \emph{Burge conditions} \cite{BURGE1993210, Bershtein:2014qma}. These reduce to the requirement
that $Y_1 \supseteq \ldots \supseteq Y_m$ (which is a plane partition of maximal height $m$) when taking the $p \to \infty$ limit
of vacuum characters at fixed $p \in \mathbb{N}$. In this limit, the central charge simplifies to
\begin{equation}
  \lim_{p \to \infty} c(\mathcal{V}(m, K; p)) = m K.
\end{equation}
We therefore identify the giant graviton coefficient as
\begin{equation}
Z_{0,m}(q_1, q_2; \vec{x})|_{q=1} = \chi_{\text{vac}}^{\mathcal{V}(m,K)}(\vec{x}) \equiv \lim_{p \to \infty} \chi_{\text{vac}}^{\mathcal{V}(m,K;p)}(\vec{x})
\end{equation}
and similar for $Z_{m,0}$ with fugacities $\lim_{q \to 1} q_{2, \alpha} = x_{\alpha}^{-1}$.
Since $q = 1$ and the ratio $q_1 / q_2$ is subject to the constraint \eqref{eq:2510311953}, the right-hand side is completely determined in terms of $\vec{x}$.
In section \ref{sec:gen_func_poincare}, we give explicit examples for this when $m = 2$ and $K = 2,3$.
Related limits of the M5-brane giant graviton coefficients in the Higgs and Coulomb branch were also explored in the recent work \cite{hayashi2025}.
The vertex algebra $\mathcal{V}(m,K; p)$ in eq.~(\ref{eq:full_algebra}) originally arises in the context of the AGT correspondence \cite{alday2010, Belavin:2011sw, Alfimov:2013cqa}.
Its appearance here seems reminiscent of the observation \cite{nishioka2011} that $m$ M5-branes on $\mathbb{C}^2 / \mathbb{Z}_K$ realise a 2d CFT with the same symmetry algebra.

\paragraph{}
When both $L$ and $K$ are arbitrary, there are $L$ wrapping numbers $\mathbf{m}_{\alpha, a}$ where $\alpha = 0, \ldots, K-1$ and $a = 1, \ldots, L_{\alpha}$.
The generating function is composed out of building blocks of the case with a single wrapping number.
In the unrefined limit $q \to 1$, these building blocks factorise, yielding
\begin{equation}
  Z_{\mathbf{0}, \mathbf{m}}(q_1, q_2; \vec{x})|_{q=1} =
  \prod_{\alpha = 0}^{K-1} \prod_{a=1}^{L_{\alpha}} \chi_{\text{vac}, \alpha}^{\mathcal{V}(\mathbf{m}_{\alpha,a}, K)}(\vec{x}).
\end{equation}
The product on the right-hand side can be thought of as the character of a direct sum of parafermionic $W$-algebras. 
The extra subscript $\alpha$ on $\chi_{\text{vac}, \alpha}$ remembers which node the flavour was originally attached to and has the effect of cyclically shifting the fugacities $x_{\beta} \to x_{\beta + \alpha}$ (this is interpreted as a $\mathbb{Z}_K$ index, $\beta \sim \beta + K$).
So far, we were only able to prove this factorisation in the plethystic case $K = 1$.
It follows naturally if the interpretation of $Z_{\mathbf{0}, \mathbf{m}}$ as $q$-counting plane partitions is correct.

\paragraph{}
The remainder of the paper is structured as follows.
In section \ref{sec:adhm_quiver}, we review the giant graviton expansion for the ADHM quiver with a single flavour, corresponding to $L = K = 1$.
We demonstrate our technique for deriving the giant graviton coefficients and show that the result coincides with the known explicit form.
As we will see, the same method also allows for evaluation of coefficients $Z_{m_1, m_2}$ with both $m_1$ and $m_2$ non-zero.
Finally, we comment on the state of deriving the giant graviton coefficients from direct localisation results for the 6d $(2,0)$ superconformal index.
While a full derivation is currently only possible in the unrefined limit $q \to 1$, we provide numeric evidence that the general case $q \not= 1$ is obtained in the same way.

In section \ref{sec:general_quiver}, we generalise the previous results to arbitrary $L$ and $K$, which includes both the Higgs and Coulomb branch Hilbert series of the ADHM quiver with $K > 1$ flavours.
While we do not attempt this, it should be possible to derive the giant graviton coefficients in a similar way from the 6d $(2,0)$ superconformal index in the presence of defects \cite{Bullimore:2014upa}.

In section \ref{sec:gen_func_poincare}, we show how to arrange our results into the fermionic formulas (\ref{eq:2509290955}, \ref{eq:2510311206}).
We comment on their relationship to characters of the vertex algebra $\mathcal{V}$ via the AGT correspondence and demonstrate explicitly with some examples how the vacuum character $\chi_{\text{vac}}^{\mathcal{V}(m,K)}$ decomposes into characters of affine Lie algebras and $W$-algebras.

\section{ADHM quiver}
\label{sec:adhm_quiver}

The world volume theory of $N$ coincidental M2-branes at the origin of
$\mathbb{C}^4$ can be
given the UV-description of ADHM theory with one flavour,
which is a 3d $\mathcal{N} = 4$ $U(N)$ quiver gauge theory containing
one adjoint hyper and $N_f = 1$ fundamental hypermultiplets.
The moduli space has two branches, the Higgs branch $\mathcal{M}_H$
and the Coulomb branch $\mathcal{M}_C$, which are related to each
other via mirror symmetry \cite{deBoer:1996mp}. ADHM with one
flavour is self-dual, Higgs and Coulomb branch being
the symmetric product of $N$ copies of $\mathbb{C}^2$,
\begin{equation}
	\label{eq:moduli_space}
	\mathcal{M}_H = \mathcal{M}_C =
	\operatorname{Sym}_N [\mathbb{C}^2].
\end{equation}
In the infrared, this theory is dual to $\text{AdS}_4 \times S^7$ and gets enhanced
to $\mathcal{N} = 8$ supersymmetry with bosonic symmetry group
\begin{equation*}
	\underset{E, J}{\operatorname{SO}(3,2)} \times \underset{Q_1, Q_2, Q_3, Q_4}{\operatorname{SO}(8)_R}.
\end{equation*}
Here, $E$, $J$ and $Q_I$, $I = 1, \dots, 4$ are the Cartan generators.
The superconformal index is defined as
\begin{equation}
	\mathcal{I}_N(y, q_I) = \operatorname{Tr}
	\left[(-1)^F y^J \prod_{I=1}^{4} q_I^{Q_I} \right],
\end{equation}
where the fugacities are subject to the condition
\begin{equation}
	\label{eq:param_constraint}
	y = q_1 q_2 q_3 q_4.
\end{equation}
This or a similar condition is necessary in order to preserve supersymmetry.

\subsection{Higgs and Coulomb branch limit}

The Higgs branch limit corresponds to $q_3, q_4 \to 0$ and
the Coulomb branch limit is $q_1, q_2 \to 0$, together with
$y \to 0$ such that the constraint~(\ref{eq:param_constraint}) is fulfilled while keeping
$q_1, q_2$ (resp. $q_3, q_4$) arbitrary.
The superconformal index reduces to the Hilbert series of the Higgs or Coulomb branch
moduli space, \cite{Razamat:2014pta}
\begin{equation}
	\lim_{q_3, q_4, y \to 0} \mathcal{I}_N(y, q_I) =
	\operatorname{Hilb}[\mathcal{M}_H](q_1, q_2) \equiv
	H_N(q_1, q_2).
\end{equation}
The Hilbert series of the abelian $N = 1$ theory is
\begin{equation}
	\operatorname{Hilb}[\mathcal{M}_H] =
	\operatorname{Hilb}[\mathbb{C}^2] = \frac{1}{1 - q_1} \frac{1}{1 - q_2}.
\end{equation}
Since the moduli space \eqref{eq:moduli_space} for $N > 1$ is a symmetric product of $N$ copies,
the generating function $H(\Lambda)$ of the Hilbert series $H_N(q_1, q_2)$ is plethystic,
\begin{equation}
  \label{eq:2510171503}
  H(\Lambda) =
	\sum_{N=0}^{\infty} \Lambda^N H_N(q_1, q_2) = \operatorname{PE}[\Lambda H_1(q_1, q_2)]
	\equiv \prod_{k_1, k_2 = 0}^{\infty} \frac{1}{1 - \Lambda q_1^{k_1} q_2^{k_2}}.
\end{equation}
We call a function plethystic if it can be written as the plethystic exponential $\operatorname{PE} [ f(x_1, x_2, \ldots) ]$ of some function $f$ with arguments $x_1, x_2, \ldots$.
The plethystic exponential is formally defined as
\begin{equation}
  \operatorname{PE} [f(x_1, x_2, \ldots)] \equiv
  \exp \sum_{i=1}^{\infty} \frac{1}{i} f(x_1^i, x_2^i, \ldots).
\end{equation}
It satisfies the rules
\begin{equation}
  \operatorname{PE} [c \, x^a] = \frac{1}{(1 - x^a)^c}
\end{equation}
and
\begin{equation}
  \operatorname{PE} [f] \operatorname{PE} [g] = \operatorname{PE} [f+g],
  \quad
  \operatorname{PE} [-f] = \operatorname{PE} [f]^{-1}.
\end{equation}

\paragraph{}
ADHM theory with one flavour, $\mathcal{T}_N[1,1]$, has a single fugacity $x$ for the $SU(2)_x$ Higgs branch isometry which is completely constrained in terms of $q_1$ and $q_2$,
\begin{equation}
  x = (q_1 / q_2)^{1/2}.
\end{equation}
Nevertheless, we will use the notation $H_N(q_1, q_2; x)$ to emphasise the dependence on $x$. We call the remaining independent fugacity $q = q_1 q_2$.

\paragraph{}
While the Higgs branch has the clearer physical interpretation since it does not receive quantum corrections,
it turns out that the Coulomb branch formulation \cite{Cremonesi:2013lqa, Cremonesi:2014xha, cremonesi2014a}
of the mirror theory is more convenient for deriving an expression for the giant graviton coefficients.
ADHM with one flavour is self-mirror and the Coulomb branch formulation expresses the Hilbert series of $\mathbb{C}^2$
in terms of monopole operators as
\begin{equation}
	\label{eq:hilbert_series}
  H_N(q_1, q_2; x) = \sum_{\mathbf{n} \in \mathbb{Z}^N / S_N}
	q^{\Delta[\mathbf{n}]} x^{|\mathbf{n}|} P_{\mathbf{n}}(q).
\end{equation}
The notation is as follows.
$\mathbf{n} = (\mathbf{n}_1, \dots, \mathbf{n}_N) \in \mathbb{Z}^N / S_N$ denotes an ordered tuple of $N$
integers, $\mathbf{n}_1 \geq \ldots \geq \mathbf{n}_N$. We set
\begin{equation}
	\label{eq:def_weight}
	|\mathbf{n}| \equiv \sum_{a=1}^N \mathbf{n}_a
\end{equation}
(without taking the absolute value of the $\mathbf{n}_a$).
$\Delta[\mathbf{n}]$ is the monopole dimension of the operator given by $\mathbf{n}$, defined by
\begin{equation}
	\Delta[\mathbf{n}] \equiv \frac{1}{2}
	\sum_{a=1}^N |\mathbf{n}_a|.
\end{equation}
These operators are dressed by the classical factor
\begin{equation}
	\label{eq:dressing}
	P_{\mathbf{n}}(q) \equiv \prod_{k \in \mathbb{Z}}
	\prod_{l=1}^{\mu_k(\mathbf{n})} \frac{1}{1 - q^{l}},
\end{equation}
where $\mu_k(\mathbf{n})$ denotes the multiplicity of $k \in \mathbb{Z}$
in $\mathbf{n} \in \mathbb{Z}^N / S_N$.

\paragraph{}
The single flavour has a $U(1)$ global symmetry which admits magnetic background charge $m$. That changes the monopole formula to
\begin{equation}
  \Delta[\mathbf{n}, m] = \frac{1}{2} \sum_{a=1}^{N} |\mathbf{n}_a - m|.
\end{equation}
However, shifting $\mathbf{n}_a \to \mathbf{n}_a + m$ in the summation shows that the only effect is an overall factor of $x^{N m}$,
which can be absorbed by introducing a fugacity for $m$. We will define the Hilbert series with magnetic background charge in
such a way that it does not depend on the overall shift symmetry,
\begin{equation}
  H_N^{[m]}(q_1, q_2; x) \equiv 
  x^{-N m}
  \sum_{\mathbf{n} \in \mathbb{Z}^N / S_N}
	q^{\Delta[\mathbf{n}, m]} x^{|\mathbf{n}|} P_{\mathbf{n}}(q) =
  H_N(q_1, q_2; x).
\end{equation}
We also define a rescaled Hilbert series as
\begin{equation}
  \label{eq:2509191006}
  \bar{H}_N^m(q_1, q_2; x) \equiv q^{-N |m| / 2} x^{N m} H_N(q_1, q_2; x) = H_N(q_1, q_2; x) \times \begin{cases}
		q_2^{- N |m|} & m > 0 \\
		q_1^{- N |m|} & m < 0
	\end{cases}
\end{equation}
which does depend on the overall value of $m$.

\subsection{Giant graviton expansion}
\label{sec:adhm_gge}

In this section, we review the derivation \cite{Gaiotto:2021xce} of the giant graviton coefficients from
the plethystic generating function. This provides explicit expressions to compare the background charge approach with.
The giant graviton expansion for $N$ M2-branes is the statement that \cite{Arai:2020uwd}
\begin{equation}
	\mathcal{I}_N(y, q_I) = \mathcal{I}_{\infty}(y, q_I)
	\sum_{n_1, n_2, n_3, n_4 = 0}^{\infty}
	\left(\prod_{I=1}^4 q_I^{N n_I}\right) Z_{n_1, n_2, n_3, n_4}(y, q_I)
\end{equation}
where $Z_{n_1, n_2, n_3, n_4}$ is the index of a 6d theory.
In the Higgs branch limit, this becomes
\begin{equation}
	\label{eq:coulomb_gge}
	H_N(q_1, q_2) = H_{\infty}(q_1, q_2)
	\sum_{n_1, n_2 = 0}^{\infty} q_1^{N n_1} q_2^{N n_2}
	Z_{n_1, n_2}(q_1, q_2)
\end{equation}
with
\begin{equation}
	\label{eq:twisted_limit}
	Z_{n_1, n_2}(q_1, q_2) \equiv \lim_{y, q_3, q_4 \to 0}
	Z_{n_1, n_2, 0, 0}(y, q_I).
\end{equation}
The Kaluza-Klein contribution can be extracted as 
\begin{equation}
	\begin{aligned}
		H_{\infty}(q_1, q_2) & = \lim_{N \to \infty} H(\Lambda)|_{\Lambda^N} =
      \lim_{N \to \infty} \oint \frac{d \Lambda}{2 \pi i \Lambda^{N+1}} H(\Lambda).
			\end{aligned}
\end{equation}
This expression has poles at $\Lambda = q_1^{-k_1} q_2^{-k_2}$.
For the index to be well-defined, we assume $|q_1|, |q_2| < 1$.
In the limit $N \to \infty$, the pre-factor $\Lambda^{-N-1}$
will suppress all poles except for $\Lambda = 1$. Since
\begin{equation}
  H(\Lambda) = \operatorname{PE}[\Lambda H_1(q_1, q_2)] = \frac{1}{1 - \Lambda} \operatorname{PE}[\Lambda (H_1(q_1, q_2) - 1)]
\end{equation}
and the latter factor is regular as $\Lambda \to 1$,
\begin{equation}
  H_{\infty}(q_1, q_2) = 
  - \operatorname{Res}_{\Lambda = 1} H(\Lambda)
    = \operatorname{PE}\left[
			\frac{1}{1-q_1} \frac{1}{1-q_2} - 1
			\right]
		= \prod_{\substack{k_1, k_2 \geq 0                                   \\(k_1, k_2) \not= (0,0)}} \frac{1}{1 - q_1^{k_1} q_2^{k_2}}.
\end{equation}
Interestingly, the giant graviton coefficients can be derived by considering the
remaining poles of the generating function $H(\Lambda)$.
The Hilbert series $H_N(q_1, q_2)$ at finite $N$ is
\begin{equation}
  \label{eq:contour_integral}
	H_N(q_1, q_2) =
	\frac{1}{2 \pi i}
	\oint \frac{d \Lambda}{\Lambda^{N+1}}
	\prod_{k_1, k_2 = 0}^{\infty} \frac{1}{1 - \Lambda q_1^{k_1} q_2^{k_2}},
\end{equation}
integrated over a circle of radius $r < 1$.
The poles are at $\Lambda = 0$ and $\Lambda = q_1^{-n_1} q_2^{-n_2}$ for $n_1, n_2 \in \mathbb{N}_0$.
Summing over all poles outside of the contour leads to precisely the form of the giant graviton
expansion, with contributions
\begin{equation}
	H_{\infty}(q_1, q_2) Z_{n_1, n_2}(q_1, q_2) =
	\prod_{\substack{k_1 \geq -n_1, k_2 \geq -n_2\\(k_1, k_2) \not= (0,0)}}
	\frac{1}{1 - q_1^{k_1} q_2^{k_2}}.
\end{equation}
In particular, this means the contributions of branes wrapped around different
cycles factorise,
\begin{equation}
  \label{eq:2510171521}
	Z_{n_1, n_2}(q_1, q_2) = \left[
    \prod_{k_1 = -n_1}^{-1} 
	\prod_{k_2 = -n_2}^{-1} \frac{1}{1 - q_1^{k_1} q_2^{k_2}}
	\right]
	Z_{n_1, 0}(q_1, q_2) Z_{0, n_2}(q_1, q_2).
\end{equation}
The factors
\begin{equation}
	\label{eq:Z_n_0}
	Z_{n_1, 0}(q_1, q_2) = \prod_{k_1 = -n_1}^{-1} \prod_{k_2=0}^{\infty} \frac{1}{1 - q_1^{k_1} q_2^{k_2}}
\end{equation}
and
\begin{equation}
  \label{eq:2510211346}
	Z_{0, n_2}(q_1, q_2) = \prod_{k_1 = 0}^{\infty} \prod_{k_2=-n_2}^{-1} \frac{1}{1 - q_1^{k_1} q_2^{k_2}}
\end{equation}
are interpreted as $n_I$ M5-branes around corresponding cycles:
The intersection term is expected to stem from M2-branes stretched between M5-branes.
Note that the factorisation \eqref{eq:2510171521} is different from the Schur-like one
proposed in \cite{Hayashi:2024aaf}.
Part of the problem is that $Z_{n_1, n_2}$ for generic $n_1, n_2$
is not analytic at $q_1 = 0$ or $q_2 = 0$. In particular, $Z_{n_1, 0}|_{q_1 = 0} = 0$ and
$Z_{0, n_2}|_{q_2 = 0} = 0$. This property can be used to reduce the giant graviton expansion to
\begin{equation}
	H_N(q_1, q_2) = H_{\infty}(q_1, q_2)
	\sum_{n_1 = 0}^{\infty} q_1^{N n_1} Z_{n_1, 0}(q_1, q_2)
\end{equation}
when expanded around $q_2 = 0$ first, and similar for when expanded
around $q_1 = 0$ first.
At these points, none of the terms $Z_{n_1, n_2}$ with $n_1 > 0$ and $n_2 > 0$ contribute.

\subsection{Relation to magnetic background charge}
\label{sec:background_flux_simple_adhm}

From the general form of giant graviton expansions as in \eqref{eq:coulomb_gge},
it can be seen that the coefficients $Z_{n_1, n_2}(q_1, q_2)$ are always related to
residues at specific poles in the grand canonical ensemble. Namely,
\begin{equation*}
  H(\Lambda) =
	\sum_{N=0}^{\infty} \Lambda^N H_N(q_1, q_2) = H_{\infty}(q_1, q_2)
	\sum_{n_1, n_2 = 0}^{\infty} \frac{1}{1- \Lambda q_1^{n_1} q_2^{n_2}}
	Z_{n_1, n_2}(q_1, q_2)
\end{equation*}
such that
\begin{equation}
  Z_{m_1, m_2} = - \frac{q_1^{m_1} q_2^{m_2}}{H_{\infty}} \operatorname{Res}_{\Lambda = q_1^{-m_1} q_2^{-m_2}} H(\Lambda).
\end{equation}
Rescaling $\Lambda \to \Lambda q_1^{-m_1} q_2^{-m_2}$,
\begin{equation}
  \label{eq:2510211348}
	Z_{m_1, m_2}(q_1, q_2) = - \frac{1}{H_{\infty}(q_1, q_2)} \operatorname{Res}_{\Lambda = 1}
  H(\Lambda q_1^{-m_1} q_2^{-m_2}).
\end{equation}
In the following, we present a trick based on the mirror Coulomb branch expression \eqref{eq:hilbert_series}
to compute this residue by factorising $H(\Lambda)$,
due to \cite{Gaiotto:2008ak, Barns-Graham:2018qvx}.
Focus first on $m_1 = 0$ or $m_2 = 0$.
In these cases the pre-factor $q_I^{-N m_I}$ combines with $H_N$ into
the rescaled Hilbert series with magnetic background charge \eqref{eq:2509191006},
\begin{align}
  q_1^{-N m_1} H_N(q_1, q_2; x) &= \bar{H}_N^{-m_1}(q_1, q_2; x) \quad \text{or} \\
  q_2^{-N m_2} H_N(q_1, q_2; x) &= \bar{H}_N^{m_2}(q_1, q_2; x).
\end{align}
In other words, we are interested in computing the residue at $\Lambda = 1$ of
\begin{equation}
  \bar{H}^m(\Lambda; q_1, q_2; x) = \sum_{N=0}^{\infty} \Lambda^N \bar{H}_N^m(q_1, q_2; x)
\end{equation}
for $m \in \mathbb{Z}$.
By shifting the summation variables, the rescaled Hilbert series with background charge can be brought into the form
\begin{equation}
  \label{eq:2510191432}
	\bar{H}_N^m(q_1, q_2; x) = q^{-N |m| / 2} \sum_{\mathbf{n} \in \mathbb{Z}^N / S_N} q^{\Delta[\mathbf{n}, m]}
	x^{|\mathbf{n}|} P_{\mathbf{n}}(q)
\end{equation}
with $P_{\mathbf{n}}(q)$ and $\Delta[\mathbf{n},m]$ as before.
We split $\mathbf{n} = (\mathbf{n}_1, \ldots, \mathbf{n}_N)$ into three lists:
$\pi_a$, $a = 1, \ldots, l_1$, denotes the positive values,
negative values are $-\nu_a$, $a = 1, \ldots, l_2$ and the remaining $N - l_1 - l_2$ entries are zeroes.
$\pi, \nu \in \mathcal{P}$ are partitions of length $l_1$ and $l_2$, respectively (we write $l(\pi) = l_1$,
$l(\nu) = l_2$).
Then,
\begin{equation}
	\begin{aligned}
		x^{|\mathbf{n}|}    & = x^{|\pi| - |\nu|} \quad \text{and}                                       \\
		P_{\mathbf{n}}(q) & = P_{\pi}(q) P_{\nu}(q) \prod_{l=1}^{N-l_1 - l_2} \frac{1}{1 - q^{l}}
	\end{aligned}
\end{equation}
where $|\lambda| \equiv \sum_{a=1}^{l(\lambda)} \lambda_a$ is the
weight of the partition $\lambda$.
To split up $q^{\Delta[\mathbf{n},m]}$, define
\begin{equation}
	\label{eq:B_min_zero}
	B(a,b) \equiv \frac{1}{2} (|a| + |b| - |a-b|) =
	\begin{cases}
		\min(|a|, |b|) & a \cdot b > 0 \\
		0              & \text{else}
	\end{cases}.
\end{equation}
The purpose of $B(a,b)$ is to split up the dependence on $|\mathbf{n}_a - m|$, such that
the $\pi_a$ and $\nu_a$ can be cleanly separated.
This necessitates separating $m$ into positive, negative and zero part as well: we define
\begin{equation}
  m^+ = \begin{cases}
    m & \text{if } m > 0 \\
    0 & \text{else}
  \end{cases}
  \quad \text{and} \quad
  m^- = \begin{cases}
    -m & \text{if } m < 0 \\
    0 & \text{else}
  \end{cases}.
\end{equation}
The scaling dimension becomes
\begin{equation}
	\begin{aligned}
		\Delta[\mathbf{n},m] & = - \sum_{a=1}^N B(\mathbf{n}_a, m) + \frac{1}{2} \sum_{a=1}^N (|\mathbf{n}_a| + |m|)    \\
		                     & \equiv h(\pi, m^+) + h(\nu, m^-) + \frac{1}{2} (|\pi| + |\nu| + N |m|)
	\end{aligned}
\end{equation}
where
\begin{equation*}
	h(\lambda, \mu) = - \sum_{a=1}^{l(\lambda)} B(\lambda_a, \mu) \quad
	\text{for} \quad \lambda \in \mathcal{P}, \mu \in \mathbb{Z}.
\end{equation*}
The last term is cancelled by the prefactor $q^{-N |m| / 2}$ in eq.~\eqref{eq:2510191432}
and the rescaled Hilbert series can be written as
\begin{equation}
	\begin{aligned}
		\bar{H}_N^m(q_1, q_2; x) = \sum_{0 \leq l_1 + l_2 \leq N}
           & \left[\sum_{\substack{\pi \in \mathcal{P}\\ l(\pi) = l_1}} q^{h(\pi, m^+)} q_1^{|\pi|} P_{\pi}(q) \right]
           \left[\sum_{\substack{\nu \in \mathcal{P}\\ l(\nu) =l_2}} q^{h(\nu, m^-)} q_2^{|\nu|} P_{\nu}(q) \right]          \\
		\times & \left[\prod_{l=1}^{N - l_1 - l_2} \frac{1}{1 - q^{l}} \right].
	\end{aligned}
\end{equation}
Defining
\begin{equation}
	\label{eq:r_l_adhm}
	\begin{aligned}
		r_{l}(m, q^{-1}, y) & \equiv
    \sum_{\substack{\pi \in \mathcal{P}\\l(\pi) = l}} q^{h(\pi, m)} y^{|\pi|} P_{\pi}(q)
		\quad \text{and}                                                            \\
		r_{l}^{(0)}(q) & \equiv P_{(0)^{l}}(q) = \prod_{k=1}^{l} \frac{1}{1 - q^k},
	\end{aligned}
\end{equation}
the above sum becomes
\begin{equation}
  \label{eq:2510261729}
	\bar{H}_N^m(q_1, q_2) = \sum_{0 \leq l_1 + l_2 \leq N}
	r_{l_1}(m^+, q^{-1}, q_1) r_{l_2}(m^-, q^{-1}, q_2) r_{N - l_1 - l_2}^{(0)} (q).
\end{equation}
Hence, the grand canonical series factorises as
\begin{equation}
  \label{eq:2510191434}
  \bar{H}^m(\Lambda; q_1, q_2) = 
	\sum_{N=0}^{\infty} \Lambda^N \bar{H}_N^m(q_1, q_2) =
	\left[\sum_{l=0}^{\infty} \Lambda^{l} r_{l}^{(+)} \right]
	\left[\sum_{l=0}^{\infty} \Lambda^{l} r_{l}^{(-)} \right]
	\left[\sum_{l=0}^{\infty} \Lambda^{l} r_{l}^{(0)} \right].
\end{equation}
In appendix \ref{sec:fermionic_forms_and_bg_flux} we show that the first two factors
in the limit $\Lambda \to 1$ become the ``Hausel generating function''
\begin{equation}
	r(m^{T}, q^{-1}, y),
\end{equation}
which is defined in the appendix. $m^{T} = (1, \ldots, 1) \equiv (1)^{m}$ refers to the transposed partition
of $m \in \mathbb{N}_0 \subset \mathcal{P}$. The last factor in \eqref{eq:2510191434}
is by the $q$-binomial theorem
\begin{equation}
	\sum_{l=0}^{\infty} \Lambda^l \prod_{k=1}^{l} \frac{1}{1 - q^{k}} =
	\prod_{k=0}^{\infty} \frac{1}{1 - \Lambda q^{k}},
\end{equation}
which has a simple pole at $\Lambda = 1$. Together,
\begin{equation}
  \label{eq:2510241533}
	- \operatorname{Res}_{\Lambda = 1} \bar{H}^m(\Lambda; q_1, q_2) =
	r(m^{+T}, q^{-1}, q_1) \, r(m^{-T}, q^{-1}, q_2) \, \prod_{k=1}^{\infty}
	\frac{1}{1 - q^{k}}.
\end{equation}
As observed before,
\begin{equation}
  H_{\infty}(q_1, q_2) = - \operatorname{Res}_{\Lambda = 1} H(\Lambda) = - \operatorname{Res}_{\Lambda = 1} \bar{H}^{0}(\Lambda; q_1, q_2),
\end{equation}
such that the giant graviton coefficients can be written as the ratios
\begin{align}
  Z_{0,m}(q_1, q_2) &= \frac{r(m^T, q^{-1}, q_1)}{r(0, q^{-1}, q_1)}
  \quad \text{and} \label{eq:2510241631}\\
  Z_{m,0}(q_1, q_2) &= \frac{r(m^T, q^{-1}, q_2)}{r(0, q^{-1}, q_2)}. \label{eq:2510241631-2}
\end{align}
The Hausel generating function $r$ can be related to fermionic forms.
We will postpone this discussion until section \ref{sec:fermionic_forms}.

\paragraph{}
There is a subtlety related to analytic continuation. In this plethystic example,
the rescaled Hilbert series (with $m > 0$) satisfies
\begin{equation}
	\begin{aligned}
		\bar{H}_N^m(q_1, q_2) = q_2^{-N m} \operatorname{PE}\left[\Lambda \frac{1}{1 - q_1} \frac{1}{1 - q_2} \right]\Bigg|_{\Lambda^N}
		= \operatorname{PE}\left[\Lambda q_2^{-m} \frac{1}{1 - q_1} \frac{1}{1 - q_2} \right]\Bigg|_{\Lambda^N}.
	\end{aligned}
\end{equation}
The residue at $\Lambda = 1$ is easily computed,
\begin{equation}
  \label{eq:2510241532}
	- \operatorname{Res}_{\Lambda = 1} \bar{H}^m(\Lambda) = \operatorname{PE} \left[
	\frac{q_2^{-m}}{(1-q_1)(1-q_2)} - 1
	\right].
\end{equation}
Terms with a negative power of $q_2$ are to be interpreted as analytically continued,
\begin{equation}
\operatorname{PE}[q_2^{-m}] = \frac{1}{1 - q_2^{-m}} = - q_2^m \operatorname{PE}[q_2^m].
\end{equation}
More generally, this applies to all coefficients $q_1^a q_2^b$ where $|q_1^a q_2^b| > 1$.
A necessary condition for $\bar{H}_{\infty}^m$ to contribute to the giant graviton expansion
is that only finitely many such terms are analytically continued \cite{Imamura:2022aua}.
For example, a formal expansion around $q_2 = 0$ first means that $|q_2| \ll |q_1^a| < 1$ for
any $a \in \mathbb{N}$. Then, there are infinitely many contributions $|q_2^{-1} q_1^a| \gg 1$
such that the giant graviton coefficient decouples. Formally expanding around $q_1 = 0$ instead
results in only finitely many contributions $q_2^{-m} + \ldots + q_2^{-1}$ that need to be
analytically continued.

\paragraph{}
In the expression \eqref{eq:2510241533} of the residue in terms of Hausel generating functions, this manifests itself
as follows.
The function $r(m^{+T}, q^{-1}, q_1)$ is inherently defined for $|q_1| \ll |q| < 1$.
But this implies $|q_2^{-1}| = |q_1/q| < 1$.
The necessity of analytic continuation makes itself most apparent when expanded in the variables $q$ and $q_2^{-1}$,
\begin{equation}
  r(m^{+T}, q^{-1}, q_1) = \sum_{n=0}^{\infty} a_n(q_2^{-1}) \, q^{n}.
\end{equation}
In order to compare this numerically to an expansion in $|q_1|, |q_2| < 1$, the functions $a_n(q_2^{-1})$ have to be approximated and analytically continued to $|q_2^{-1}| > 1$.
To make matters worse, the second factor $r(m^{-T}, q^{-1}, q_2)$ is defined conversely for $|q_2| \ll |q| < 1$, which has a different regime of validity as the term with $m^{+T}$.
In the giant graviton coefficients $Z_{0,m}$ and $Z_{m,0}$, which involve a ratio of Hausel functions, only the first problem about analytic continuation of $q_2^{-1}$ (or $q_1^{-1}$) is a concern.
However, the mixed coefficients $Z_{m_1, m_2}$ which we compute in the next section contain factors of both $Z_{0,m_2}$ and $Z_{m_1, 0}$ and require both parts to be separately analytically continued in order to be a valid expression in any regime.

\paragraph{}
The plethystic expression \eqref{eq:2510241532} can be explicitly identified
with \eqref{eq:2510241533} as follows.
\begin{equation}
  \begin{aligned}
    - \operatorname{Res}_{\Lambda = 1} \bar{H}^m(\Lambda)
    =&
  \underbrace{\operatorname{PE} \left[\frac{1}{1 - q_1} \frac{q_2^{-m} - 1}{1 - q_2} + \frac{q_1}{1 - q_1} \frac{1}{1 - q}\right]}_{= r(m^T, q^{-1}, q_1)} \\
     &\times
  \underbrace{\operatorname{PE} \left[\frac{q_2}{1 - q_2} \frac{1}{1 - q}\right]}_{= r(0, q^{-1}, q_2)}
     \times
     \underbrace{\operatorname{PE} \left[\frac{q}{1 - q}\right]}_{= \prod_{k=1}^{\infty} \frac{1}{1 - q^{k}}}.
  \end{aligned}
\end{equation}
In a slight abuse of notation, we will sometimes refer to the residue at $\Lambda = 1$ as the $N \to \infty$ limit $\bar{H}_{\infty}^m(q_1, q_2)$.
This is justified in the sense that \eqref{eq:2510241533} is the $N \to \infty$ limit of \eqref{eq:2510261729}.
While $\bar{H}_{\infty}^m(q_1, q_2)$ as a series expansion in $q$ and $x$ diverges due to poles at $|\Lambda| \sim |q_1^a q_2^b| < 1$, the residue at $\Lambda = 1$, which is obtained by analytically continuing the three factors individually, is finite.
Then, the coefficient $Z_{0,m}$ is the ratio
\begin{equation}
	Z_{0,m}(q_1, q_2) =
	\frac{\bar{H}_{\infty}^m(q_1, q_2)}{H_{\infty}(q_1, q_2)} =
	\operatorname{PE} \left[
    \frac{1}{1-q_1}
	\frac{q_2^{-m} - 1}{1-q_2}
	\right] =
	\prod_{k_1 = 0}^{\infty} \prod_{k_2 = -m}^{-1}
	\frac{1}{1 - q_1^{k_1} q_2^{k_2}}
\end{equation}
and similar for $Z_{m,0}$, in agreement with \eqref{eq:Z_n_0}
and \eqref{eq:2510211346}.

\subsection{Coefficients $Z_{m_1, m_2}$ with two non-zero wrapping numbers}

In this subsection, we briefly comment on the general case \eqref{eq:2510211348}
with two non-zero winding numbers $m_1, m_2 \not= 0$. We will assume $m_2 \geq m_1$. In terms of variables $q$ and $x$,
\begin{equation}
  \begin{aligned}
    H(\Lambda q_1^{-m_1} q_2^{-m_2})
    &= \sum_{N=0}^{\infty} \Lambda^N q^{-N (m_1 + m_2) / 2} x^{N (m_2 - m_1)} H_N(q_1, q_2; x) \\
    &= \sum_{N=0}^{\infty} \Lambda^N q^{-N m_1} \bar{H}_N^{m_2 - m_1}(q_1, q_2, x).
  \end{aligned}
\end{equation}
As before, this splits into
\begin{equation}
  \label{eq:2510211403}
  H(\Lambda q_1^{-m_1} q_2^{-m_2}) = 
	\left[\sum_{l=0}^{\infty} \Lambda^{l} q^{- l m_1} r_{l}^{(+)} \right]
	\left[\sum_{l=0}^{\infty} \Lambda^{l} q^{- l m_1} r_{l}^{(-)} \right]
	\left[\sum_{l=0}^{\infty} \Lambda^{l} q^{- l m_1} r_{l}^{(0)} \right].
\end{equation}
It is not difficult to see that the first two terms, as $\Lambda \to 1$, go to
\begin{equation}
  r(\Delta m^T + m_1, q^{-1}, q_1) \,
  r(m_1, q^{-1}, q_2),
\end{equation}
where $\Delta m = m_2 - m_1$. The notation $\Delta m^T + m_1$ means adding $m_1 \in \mathbb{N}$ to the first part
in the partition $\Delta m^T = (1, \ldots, 1) = (1)^{\Delta m}$. In other words,
\begin{equation}
  \Delta m^T + m_1 = (1 + m_1, \underbrace{1, \ldots, 1}_{\Delta m - 1}).
\end{equation}
The third term in \eqref{eq:2510211403} is
\begin{equation}
  \prod_{k=1}^{\infty} \frac{1}{1 - \Lambda q^{-m_1 + k}}
\end{equation}
and has again a simple pole at $\Lambda = 1$, such that in total
\begin{equation}
  Z_{m_1, m_2} = \frac{r(\Delta m^T + m_1, q^{-1}, q_1)}{r(0, q^{-1}, q_1)}
  \frac{r(m_1, q^{-1}, q_2)}{r(0, q^{-1}, q_2)}
  \prod_{k = -m_1}^{-1} \frac{1}{1 - q^{k}}
  \quad \text{if} \quad
  m_2 \geq m_1
\end{equation}
and similarly
\begin{equation}
  Z_{m_1, m_2} = \frac{r(m_2, q^{-1}, q_1)}{r(0, q^{-1}, q_1)}
  \frac{r((- \Delta m)^T + m_2, q^{-1}, q_2)}{r(0, q^{-1}, q_2)}
  \prod_{k = -m_2}^{-1} \frac{1}{1 - q^{k}}
  \quad \text{if} \quad
  m_1 \geq m_2.
\end{equation}
This result is consistent with eq.~\eqref{eq:2510171521}.
We will not make use of $Z_{m_1, m_2}$ in the rest of this paper.

\subsection{Twisted limit of 6d index}
\label{sec:twisted_limit}

The coefficients in the giant graviton expansion are interpreted
as the superconformal index of the world-volume theories of M5-brane configurations.
We leave those coefficients corresponding to two sets of intersecting
M5-branes for future work and instead focus on the case of a single wrapping number
$n_1 \equiv m$.\footnote{Due to the simple sum expansion,
	these terms are in principle sufficient to interpret the full Hilbert series as
  composed of M5-brane configurations of different wrapping numbers.
  }
In this case, $Z_{m,0,0,0}(y, q_I)$ is expected to correspond to
the 6d $(2,0)$ $U(m)$ superconformal index \cite{Kim:2013nva, Arai:2020uwd},
\begin{equation}
	\label{eq:6d_sc_index}
  Z_{S^5 \times S^1}^{U(m)}(\tilde{y}_i, \tilde{q}_I) = \operatorname{Tr} \left[
		(-1)^F \prod_{i=1}^{3} \tilde{y}_i^{J_i} \prod_{I=1}^{2} \tilde{q}_I^{Q_I}
		\right].
\end{equation}
The fugacities $\tilde{y}_i, \tilde{q}_I$ are related to
$y, q_I$ by \cite{Choi:2022ovw}
\begin{equation}
	\begin{aligned}
		y         & = \tilde{q}_1, \\
		q_1       & = \tilde{q}_2^{-1} \quad \text{and} \\
		q_{2,3,4} & = \tilde{y}_{1,2,3}.
	\end{aligned}
\end{equation}
This identification ensures the new fugacity constraint: eq.~\eqref{eq:param_constraint} becomes
\begin{equation}
  \tilde{q}_1 \tilde{q}_2 = \tilde{y}_1 \tilde{y}_2 \tilde{y}_3.
\end{equation}
Note that $q_1$ gets mapped to $\tilde{q}_2^{-1}$, which means that the giant graviton coefficient is
related to the superconformal index by analytic continuation.

\paragraph{}
The effect of the Higgs branch limit on the 6d superconformal index is what is called the twisted limit in \cite{Hayashi:2024aaf}.
This means $\tilde{y}_2, \tilde{y}_3, \tilde{q}_1 \to 0$ with $\tilde{y}_2 \tilde{y}_3 / \tilde{q}_1$ held fixed.
In this section we provide evidence that the giant graviton coefficients in the Higgs branch limit are indeed equal to the
twisted limit of the 6d $(2,0)$ superconformal index.

First, the abelian $m = 1$ index describing the world-volume theory of a single M5-brane is given by \cite{Bhattacharya:2008zy}
\begin{equation}
	Z_{S^5 \times S^1}^{U(1)} = \operatorname{PE} \left[
	\frac{\tilde{q}_1 + \tilde{q}_2 -
	\tilde{q}_1 \tilde{q}_2 (\tilde{y}_1^{-1}+\tilde{y}_2^{-1}+\tilde{y}_3^{-1}-1)}{
	(1-\tilde{y}_1)(1-\tilde{y}_2)(1-\tilde{y}_3)
	}
	\right].
\end{equation}
This directly reduces to
\label{eq:2510291445}
\begin{equation}
	\lim_{\substack{\tilde{y}_2, \tilde{y}_3, \tilde{q}_1 \to 0\\
			\tilde{y}_2 \tilde{y}_3 / \tilde{q}_1 = \text{const}}}
	Z_{S^5 \times S^1}^{U(1)}(\tilde{y}_i, \tilde{q}_I) =
  \operatorname{PE} \left[\frac{\tilde{q}_2}{1 - \tilde{y}_1} \right] =
	\prod_{k=0}^{\infty} \frac{1}{1 - q_1^{-1} q_2^k} = Z_{1,0}(q_1, q_2).
\end{equation}
For $m > 1$ it seems challenging to derive this directly from the localisation results \cite{Kim:2012qf, Kim:2013nva}
for the 6d superconformal index.
Part of the reason is that the $S^5$ formulation heavily involves modular transformations while
in the $\mathbb{CP}^2 \times S^1$ formulation the limit pinches the contour.
However, specialising to the unrefined case $q_1 q_2 = 1$ (which means $\tilde{y}_1 = \tilde{q}_2$),
it is easy to explicitly compare with computations in e.g. \cite{Kim:2013nva} that
\begin{equation}
  \label{eq:adhm_gen_poincare}
	Z_{S^5 \times S^1}^{U(m)}|_{q = 1} =
	\prod_{k=0}^{\infty} \prod_{a=1}^m \frac{1}{1 - q_2^{a+k}} = Z_{m,0}(q_1, q_2)|_{q = 1}.
\end{equation}
Supersymmetry dictates this, even before taking the twisted limit
on the left hand side. The equality in the unrefined limit was previously noted in \cite{Hayashi:2024aaf}.
In appendix \ref{sec:6d_sc_index}, we review the $\mathbb{CP}^2 \times S^1$ formulation of
the superconformal index, which essentially consists of a contour integral over 
three Nekrasov partition functions. These receive a perturbative and an instanton contribution,
\begin{equation}
  Z_{\text{Nek}}^{(\alpha)} = Z_{\text{pert}}^{(\alpha)} Z_{\text{inst}}^{(\alpha)},
  \quad \alpha = 1,2,3.
\end{equation}
We note that in the derivation of the unrefined limit,
one of the three instanton contributions reduces to
\begin{equation}
  \label{eq:adhm_poincare}
  \prod_{k=0}^{\infty} \frac{1}{(1 - q_2^{1+k})^m}
\end{equation}
and the other two trivialise.
Conversely, one perturbative factor trivialises while the other two remain,
but get simplified.
It is the contour integral over this residual perturbative part that cancels the superfluous factors in
\eqref{eq:adhm_poincare} and reduces it to \eqref{eq:adhm_gen_poincare}.

In appendix \ref{sec:6d_sc_index} we verify perturbatively in $q = q_1 q_2$ and $q_2$, up to
$\mathcal{O}(q_2^4)$ and for arbitrary $m$,
that the giant graviton coefficients correspond to the twisted limit of the 6d superconformal index,
even in the general case when $q \not= 1$.

\section{\texorpdfstring{$U(N)^L$}{U(N)\textasciicircum L} gauge theory with $K$ flavours}
\label{sec:general_quiver}

We move on to the general class of theories $\mathcal{T} = \mathcal{T}_N[K,L]$ with quiver diagram as in Figure~\ref{fig:quiver_diagram}.
The infrared fixed point describes the dynamics of $N$ M2-branes
probing a Calabi-Yau $\mathbb{C}^2 / \mathbb{Z}_L \times \mathbb{C}^2 / \mathbb{Z}_K$ singularity.
For $L = 1$, this reduces to the ADHM quiver with $K = K_0$ flavours.
The gravity dual at large $N$ is M-theory on $\text{AdS}_4 \times S^7 /\Gamma$ where $S^7 / \Gamma$ is the base  of the $\mathbb{C}^2 / \mathbb{Z}_L \times \mathbb{C}^2 / \mathbb{Z}_K$ cone, with $\Gamma$ acting as
\begin{equation}
  \begin{aligned}
    (z_1, z_2, z_3, z_4) &\sim (\omega_L z_1, \omega_L^{-1} z_2, z_3, z_4), \\
    (z_1, z_2, z_3, z_4) &\sim (z_1, z_2, \omega_K z_3, \omega_K^{-1} z_4).
  \end{aligned}
\end{equation}
$\omega_n = e^{2 \pi i /n}$ is the $n$-th root of unity.

\paragraph{}
The superconformal index is the function
\begin{equation}
 \mathcal{I}_{N}(y, q_I; \vec{x}, \vec{y}).
\end{equation}
Next to the angular momentum $y$, $\mathcal{I}_{N}$ depends on $L + K + 2$ fugacities.
There are four mesonic fugacities $q_1, q_2, q_3$ and $q_4$ as well as $L + K - 2$ baryonic symmetries.
We parametrise these by $K$ variables $x_{\alpha}$, $\alpha = 0, \ldots, K-1$ and $L$ variables $y_A$, $A = 0, \ldots, L-1$, subject to the constraints
\begin{equation}
  (q_1 / q_2)^{L/2} = \prod_{\alpha = 0}^{K-1} x_{\alpha} \equiv x
  \quad \text{and} \quad
  (q_3 / q_4)^{K/2} = \prod_{A = 0}^{L-1} y_{A} \equiv y.
\end{equation}
The $x_{\alpha}$ are the fugacities of the global $SU(2)_x \times SU(K)$ symmetry which are the isometries of the Higgs branch, while the $y_A$ are fugacities for the topological $U(1)^L$ symmetry, which is enhanced by quantum effects to the full $SU(2)_y \times SU(L)$ isometry group of the Coulomb branch.
We denote the remaining independent fugacity of the $R$-symmetry by
\begin{equation}
  q = q_1 q_2.
\end{equation}
One can introduce background flux for both the global and topological symmetries.
We assume that background flux for the global symmetry has the effect of shifting the $K$ flavours around the nodes and will henceforth consider the more general circular quiver gauge theory in Figure~\ref{fig:quiver_diagram_general_flavours} whilst ignoring this type of background flux.
We will comment on this more in future work.
We refer to the remaining background flux for the topological symmetry as ``baryon number''.

\subsection{Higgs and Coulomb branch limit}

The Higgs branch limit sends $y, q_3, q_4 \to 0$ such that $q_1, q_2 = \text{const}$.
The Coulomb branch isometries $y_A$ act trivially on the Higgs branch.
The remaining function
$$
H_N(q_1, q_2; \vec{x}) = \lim_{\text{Higgs}} \mathcal{I}_N(y, q_I; \vec{x}, \vec{y})
$$
is the Hilbert series of the Higgs branch moduli space.

\begin{figure}[]
	\centering
  \includegraphics{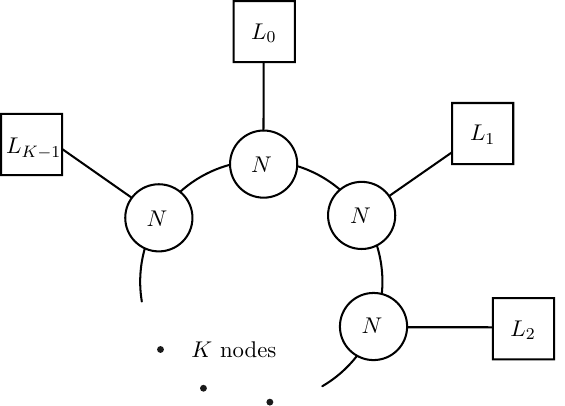}
  \caption{The quiver diagram of the mirror theory to $\mathcal{T} = \mathcal{T}_N[K,L]$,
    which is the theory $\mathcal{T}' = \mathcal{T}_N[L,K]$ with $K$ nodes and $L$ total flavours.
    The flavours are distributed over the nodes, satisfying $\sum_{\alpha = 0}^{K-1} L_{\alpha} = L$.
    The precise relation between the $L_{\alpha}$ and $K_A$ was mentioned in the introduction
    and can be found in \cite{deBoer:1996mp}.
		\label{fig:quiver_diagram_mirror}}
\end{figure}

\paragraph{}
For computational purposes, we identify the Higgs branch Hilbert series with the Coulomb branch Hilbert series of the mirror theory and leverage the monopole formula \cite{Cremonesi:2013lqa, cremonesi2014a, Cremonesi:2014xha}.
The mirror theory to $\mathcal{T} = \mathcal{T}_N[K,L]$ is $\mathcal{T}' = \mathcal{T}_N[L,K]$ with $L$ and $K$ exchanged as in Figure~\ref{fig:quiver_diagram_mirror}.
Mirror symmetry swaps particles (charged under global symmetry) and vortices (charged under topological symmetry), which means that baryon number in $\mathcal{T}$ becomes magnetic background flux $\mathbf{M} \in \mathbb{Z}^L$ for the global flavour symmetry in $\mathcal{T}'$.
We denote entries in $\mathbf{M}$ by a double index $(\alpha, a)$, where $\alpha = 0, \ldots, K-1$ and $a = 1, \ldots, L_{\alpha}$.
For a generic profile of baryonic charges $\mathbf{M}$, the Weyl group action permutes the $\mathbf{M}_{\alpha,a}$ for fixed $\alpha$ such that we may assume
\begin{equation}
 \mathbf{M}_{\alpha,1} \geq \mathbf{M}_{\alpha,2} \geq \ldots \geq \mathbf{M}_{\alpha,L_{\alpha}}
 \quad \text{for all} \quad \alpha = 0, \ldots, K-1.
\end{equation}
An overall $U(1)$ related to the shift symmetry $\mathbf{M}_{\alpha,a} \to \mathbf{M}_{\alpha,a} + m$ for $m \in \mathbb{Z}$
decouples.
We set its fugacity to one such that the Hilbert series $H_N^{[\mathbf{M}]}(q_1, q_2; x_{\alpha})$ depends only on the equivalence class $[\mathbf{M}]$ of all $\mathbf{M}_{\alpha,a}$ related by shift symmetry.
The monopole formula states that the Hilbert series is given by
\begin{equation}
  \label{eq:2509251529}
  H_{N}^{[\mathbf{M}]}(q_1, q_2; \vec{x}) =
  x^{- N |\mathbf{M}|/L}
	\sum_{\mathbf{n}^\alpha \in \mathbb{Z}^{N} / S_{N}}
	q^{\Delta[\mathbf{n}, \mathbf{M}]}
	\prod_{\alpha = 0}^{K-1}
	x_{\alpha}^{|\mathbf{n}^{\alpha}|} P_{\mathbf{n}^{\alpha}} (q),
\end{equation}
again with
\begin{equation}
  |\mathbf{n}^{\alpha}| \equiv \sum_{a=1}^{N} \mathbf{n}_a^{\alpha}
  \quad \text{and} \quad
  P_{\mathbf{n}^{\alpha}}(q) = \prod_{k \in \mathbb{Z}}^{} \prod_{l=1}^{\mu_k(\mathbf{n})} \frac{1}{1-q^{l}},
\end{equation}
as well as
\begin{equation}
  |\mathbf{M}| \equiv
  \sum_{\alpha = 0}^{K-1} \sum_{a=1}^{L_{\alpha}} \mathbf{M}_{\alpha,a}.
\end{equation}
The monopole charge is
\begin{equation}
	\Delta[\mathbf{n}, \mathbf{M}]
	=
  - \frac{1}{4} \sum_{\alpha, \beta = 0}^{K-1} C_{\alpha \beta} \sum\limits_{a,b = 1}^{N} |\mathbf{n}^{\alpha}_a
	- \mathbf{n}^{\beta}_{b}|
	+
	\frac{1}{2}
	\sum_{\alpha = 0}^{K-1}
	\sum\limits_{a = 1}^{N}
	\sum\limits_{b = 1}^{L_{\alpha}}
	|\mathbf{n}^{\alpha}_{a} - \mathbf{M}_{\alpha,b}|.
\end{equation}
Here, $C_{\alpha \beta}$ is the adjacency matrix of the gauge nodes in the graph, which coincides with the
$K \times K$ generalised Cartan matrix of $\widehat{\mathfrak{su}}(K)$.
Its components are
\begin{equation}
  \label{eq:2511031252}
C_{\alpha \beta} = 2 \delta_{\alpha \beta} - \delta_{\alpha, \beta + 1} - \delta_{\alpha, \beta - 1}
\end{equation}
 for $\alpha, \beta = 0, \ldots, K-1$, where the subscripts of the Kronecker deltas are identified modulo $K$.

\paragraph{}
It is helpful to define a rescaled Hilbert series which couples to the shift symmetry as
\begin{equation}
  \bar{H}_{N}^{\mathbf{M}}(q_1, q_2; \vec{x}) =
  q^{-N \|\mathbf{M}\| / 2} x^{N |\mathbf{M}| / L} \cdot
  H_{N}^{[\mathbf{M}]}(q_1, q_2; \vec{x})
\end{equation}
with $|\mathbf{M}|$ as before and
\begin{equation}
  \| \mathbf{M} \| = \sum_{\alpha = 0}^{K-1} \sum_{a=1}^{L_{\alpha}} |\mathbf{M}_{\alpha,a}|. 
\end{equation}
Then,
\begin{equation}
	\bar{H}_N^{\mathbf{M}}(q_1, q_2; \vec{x}) =
  H_{N}^{[\mathbf{M}]}(q_1, q_2; \vec{x}) \times \begin{cases}
    q_{2}^{-N |\mathbf{M}|} & |\mathbf{M}| > 0 \\
  q_{1}^{+ N |\mathbf{M}|} & |\mathbf{M}| < 0
	\end{cases}.
\end{equation}

\subsection{Giant graviton expansion}

In this section, we establish the form of the giant graviton expansion for general $L$ and $K$.
Except for the plethystic cases, we postpone the evaluation of the giant graviton coefficients to section \ref{sec:2510311411}.

\paragraph{$K = 1$, $L$ arbitrary:}
First, consider the giant graviton expansion for $K = 1$, but arbitrary $L \geq 1$.
The Higgs branch moduli space is $\mathcal{M}_N[1,L] = \operatorname{Sym}_N \mathbb{C}^2 / \mathbb{Z}_L$.
The Hilbert series of $\mathcal{M}_N[1,L]$ depends on the variables $q_1$ and $q_2$.
At gauge rank $N = 1$, the Hilbert series is
\begin{equation}
  H_1(q_1, q_2) = \frac{1}{1 - q_1^L} \frac{1}{1 - q_2^L}
  \frac{1 - q^L}{1 - q}.
\end{equation}
For $N > 1$, $H_N$ can be extracted from the plethystic generating function
\begin{equation}
  \sum_{N = 0}^{\infty} \Lambda^N H_N(q_1, q_2) =
  \operatorname{PE}[\Lambda H_1(q_1, q_2)] =
  \prod_{k=0}^{L-1}\prod_{k_1, k_2 = 0}^{\infty} \frac{1}{1 - \Lambda q_1^{L k_1 + k} q_2^{L k_2 + k}}.
\end{equation}
This discussion mirrors the $L = K = 1$ case.
A contour integral as in \eqref{eq:contour_integral} and summing over residues apart from zero leads to the giant graviton expansion
\begin{equation}
  H_N(q_1, q_2) = H_{\infty}(q_1, q_2) \sum_{n=0}^{L-1}
  \sum_{n_1, n_2 = 0}^{\infty} q_1^{N(L n_1 + n)} q_2^{N (L n_2 + n)}
  Z_{n, n_1, n_2}(q_1, q_2)
\end{equation}
with
\begin{equation}
  H_{\infty}(q_1, q_2) = \prod_{k = 0}^{L-1}
  \prod_{\substack{k_1, k_2 = 0 \\ (k, k_1, k_2) \not= (0,0,0)}}^{\infty}
  \frac{1}{1 - q_1^{L k_1 + k} q_2^{L k_2 + k}}
\end{equation}
and
\begin{equation}
  H_{\infty}(q_1, q_2) Z_{n, n_1, n_2}(q_1, q_2) =
 \prod_{k = 0}^{L-1}
  \prod_{\substack{k_1 = -n_1, k_2 = -n_2 \\ (k, k_1, k_2) \not= (n,0,0)}}^{\infty}
  \frac{1}{1 - q_1^{L k_1 + (k - n)} q_2^{L k_2 + (k-n)}}.
\end{equation}
The supergravity modes $H_{\infty}(q_1, q_2)$ may also be written as
\begin{equation}
  H_{\infty}(q_1, q_2) = \operatorname{PE}
  \left[
  \frac{1}{1-q_1^L} \frac{1}{1-q_2^L} \frac{1-q^L}{1 - q}
  -1
  \right].
\end{equation}
This is the result that one obtains from $L = 1$ by performing an orbifold projection onto the $\mathbb{Z}_L$-invariant part,
\begin{equation}
  H_{\infty}^{(L)}(q_1, q_2) = 
  \frac{1}{L}
  \sum_{A=0}^{L-1} H_{\infty}^{(L=1)}(\omega_L^A q_1, \omega_L^{-A} q_2).
\end{equation}
As before, wall crossing phenomena when expanding about $q_1 = 0$ or $q_2 = 0$ eliminate most of the terms $Z_{n, n_1, n_2}$ in the giant graviton expansion, such that there is a simple-sum expansion of the form
\begin{equation}
  H_{N}(q_1, q_2) = H_{\infty}(q_1, q_2)
  \sum_{n_1 = 0}^{\infty} q_1^{N L n_1} Z_{0, n_1, 0}(q_1, q_2)
\end{equation}
when expanded in $q_2$ first, with coefficients
\begin{equation}
  \label{eq:2511032151}
  Z_{0, n_1, 0}(q_1, q_2) = \prod_{k=0}^{L-1} \prod_{k_1 = -n_1}^{-1} \prod_{k_2 = 0}^{\infty} \frac{1}{1 - q_1^{L k_1 + k} q_2^{L k_2 + k}},
\end{equation}
and vice versa when expanded in $q_1$ first.
From now on, for coefficients $Z_{n,n_1, n_2}$ with $n = 0$ we will write $Z_{n_1, n_2}$ for short.
Since expansions both around $q_1 = 0$ and $q_2 = 0$ eliminate any terms with $n \not= 0$, we will ignore this distinction.

\paragraph{}
For $L > 1$, the Hilbert series also admits non-trivial baryonic charge $\mathbf{M} \in \mathbb{Z}^L$.
The effect of introducing such charge was investigated for example in \cite{Fujiwara:2023bdc} in the context of D3-brane giant gravitons on orbifolds.
In analogy to this, we expect that the effect of baryonic charge is to modify the wrapping numbers.
As an example, the simple sum expansion around $q_1 = 0$ may be written as
\begin{equation}
  H_N(q_1, q_2) = H_{\infty}(q_1, q_2) \sum_{\mathbf{n}_1 = \ldots = \mathbf{n}_L \equiv n = 0}^{\infty} q_2^{N |\mathbf{n}|} Z_{0,n}(q_1, q_2),
\end{equation}
where
\begin{equation}
  |\mathbf{n}| = \sum_{a=1}^{L} \mathbf{n}_a.
\end{equation}
The configuration vectors $\mathbf{n} \in \mathbb{N}_0^L$ are related to $\mathbf{M} = \mathbf{0} \in \mathbb{Z}^L$ by shift symmetry.
In the presence of non-trivial background flux $\mathbf{M} \not= \mathbf{0}$, the summation is again over those vectors $\mathbf{n} \in [\mathbf{M}]$ in the equivalence class of shift symmetries of $\mathbf{M}$ that only have non-negative components.
We denote this as $[\mathbf{M}]_+ \equiv [\mathbf{M}] \cap \mathbb{N}_0^L$.
The simple-sum giant graviton expansion (expanded around $q_1 = 0$) is then
\begin{equation}
  H_N^{[\mathbf{M}]}(q_1, q_2) = H_{\infty}(q_1, q_2) \sum_{\mathbf{n} \in [\mathbf{M}]_+}
  q_2^{N |\mathbf{n}|} Z_{\mathbf{0}, \mathbf{n}}(q_1, q_2).
\end{equation}
Alternatively, when expanded around $q_2 = 0$,
\begin{equation}
  H_N^{[\mathbf{M}]}(q_1, q_2) = H_{\infty}(q_1, q_2)
  \sum_{\mathbf{n} \in [-\mathbf{M}]_+}
  q_1^{N |\mathbf{n}|} Z_{\mathbf{n}, \mathbf{0}}(q_1, q_2).
\end{equation}
One important thing to note about these expansions is that the first coefficient is not trivial anymore. For example, if we assume
that $\mathbf{M}$ is maximally shifted such that all $\mathbf{M}_{a} \geq 0$ and $\mathbf{M}_L = 0$ (remember that the Weyl group symmetry is eliminated by choosing $\mathbf{M}_1 \geq \ldots \geq \mathbf{M}_L$), the first coefficient when expanded around $q_1 = 0$ is
\begin{equation}
  q_2^{N |\mathbf{M}|}  Z_{\mathbf{0}, \mathbf{M}}(q_1, q_2) \not= 1.
\end{equation}
For the expansion around $q_2 = 0$, $-\mathbf{M}$ has to be shifted and reordered. The result is
\begin{equation}
  \tilde{\mathbf{M}}_a = - \mathbf{M}_{L+1-a} + \mathbf{M}_1,
\end{equation}
satisfying $\tilde{\mathbf{M}}_1 \geq \ldots \geq \tilde{\mathbf{M}}_L = 0$. Then, the first coefficient is
\begin{equation}
  q_1^{N |\tilde{\mathbf{M}}|} Z_{\tilde{\mathbf{M}},0}(q_1, q_2) \not= 1.
\end{equation}

\paragraph{$K$ arbitrary, $L = 1$:}
We now move on to the giant graviton expansion of the Higgs branch Hilbert series of ADHM theory with $K$ flavours.
The Higgs branch moduli space $\mathcal{M}_N[K,1]$ is the moduli space of $N$ $SU(K)$ instantons on $\mathbb{C}^2$.
This is the first example that is not plethystic and for which we will need the Coulomb branch technology.
Since $L = 1$, there is no non-trivial background flux apart from shift symmetry for the topological symmetry.
The Hilbert series $H_N(q_1, q_2; \vec{x})$ depends on fugacities $q_1$, $q_2$ and $x_{\alpha}$, $\alpha = 0, \ldots, K-1$,
subject to the constraint
\begin{equation}
  (q_1 / q_2)^{1/2} = \prod_{\alpha = 0}^{K-1} x_{\alpha} \equiv x.
\end{equation}
We assume a giant graviton expansion of the form
\begin{equation}
\label{eq:2509151001}
H_{N}(q_1, q_2; \vec{x}) = H_{\infty}(q_1, q_2; \vec{x})
\sum_{n_1, n_2 = 0}^{\infty} q_1^{N n_1} q_2^{N n_2} Z_{n_1, n_2}(q_1, q_2; \vec{x}).
\end{equation}
The simple-sum truncation around $q_1 = 0$ is
\begin{equation}
  H_N(q_1, q_2; \vec{x}) = H_{\infty}(q_1, q_2; \vec{x}) \sum_{n_2 = 0}^{\infty} q_2^{N n_2} Z_{0, n_2}(q_1, q_2; \vec{x})
\end{equation}
and likewise around $q_2 = 0$.
The large $N$ limit of the Hilbert series is \cite{crew2021}
\begin{equation}
  H_{\infty}(q_1, q_2; \vec{x}) =
 \prod_{l=1}^{\infty} \frac{1}{1 - q_1^l} \frac{1}{1 - q_2^l}
\prod_{l_1, l_2 = 1}^{\infty} \prod_{\alpha, \beta =1}^{K}
\frac{1}{1 - q_1^{l_1} q_2^{l_2} z_{\alpha} / z_{\beta}}.
\end{equation}
Remember that the $x_{\alpha}$ for $\alpha = 1, \ldots, K-1$ are fugacities corresponding to the simple roots of the $\mathfrak{su}(K)$ global symmetry.
The $z_{\alpha}$ for $\alpha = 1, \ldots, K$ are a different choice of basis for $\mathfrak{su}(K)$, related to $x_{\alpha}$ by
\begin{equation}
    x_{\alpha} = z_{\alpha} / z_{\alpha+1}, \quad \alpha = 1, \ldots, K-1
  \quad \text{and} \quad
    \prod_{\alpha = 1}^{K} z_{\alpha} = 1.
\end{equation}

\paragraph{$K$ and $L$ arbitrary:}
Finally, we look at the general $U(N)^L$ quiver with $K > 1$ flavours.
The $K$ flavours are distributed onto the $L$ nodes as $K_A$, $A = 0, \ldots, L-1$ such that $\sum_{A = 0}^{L-1} K_A = K$.
The mirror quiver has $K$ nodes with $L_{\alpha}$ flavours attached to the $\alpha$-th gauge node.
The $L_{\alpha}$ are obtained from $K_A$ as in \cite{deBoer:1996mp}.
We organise the $L$ baryonic charges by double indices $(\alpha, a)$, $\mathbf{M} = \mathbf{M}_{\alpha, a}$, where $\alpha = 0, \ldots, K-1$, $a = 1, \ldots, L_{\alpha}$.
Generalising from the two previous cases, we expect a simple-sum expansion (about $q_1 = 0$) in the presence of background flux $\mathbf{M}$ of the form
\begin{equation}
  \label{eq:2509232110}
  H_{N}^{[\mathbf{M}]}(q_1, q_2; \vec{x})
          = H_{\infty}(q_1, q_2; \vec{x})
            \sum_{\mathbf{n} \in [\mathbf{M}]_+}
      q_2^{N |\mathbf{n}|} Z_{\mathbf{0},\mathbf{n}}(q_1, q_2; \vec{x}).
\end{equation}
The sum is over all $L$-tuples of non-negative integers $\mathbf{n} = (\mathbf{n}_{\alpha,a}) \in \mathbb{N}_0^{L}$ which satisfy $\mathbf{n} = \mathbf{M} + m$ for $m \in \mathbb{Z}$.
$|\mathbf{n}|$ is defined as
\begin{equation}
  |\mathbf{n}| \equiv \sum_{\alpha = 0}^{K-1} \sum_{a = 1}^{L_{\alpha}} \mathbf{n}_{\alpha,a}.
\end{equation}
When expanded about $q_2 = 0$ instead, \eqref{eq:2509232110} becomes
\begin{equation}
	\label{eq:coulomb_gge_general}
  H_{N}^{[\mathbf{M}]}(q_1, q_2; \vec{x})
          = H_{\infty}(q_1, q_2; \vec{x})
            \sum_{\mathbf{n} \in [-\mathbf{M}]_+}
      q_1^{N |\mathbf{n}|} Z_{\mathbf{n},\mathbf{0}}(q_1, q_2; \vec{x}).
\end{equation}

\subsection{Determination of giant graviton coefficients}
\label{sec:2510311411}

In the previous section we established the expected form of the giant graviton expansion for both $L$ and $K$ completely general, as well as baryonic charges $\mathbf{M}$ for the topological $U(1)^L$ symmetry.
Now, we show how this expression can be rearranged to obtain an explicit formula for the coefficients $Z_{\mathbf{n}_1,\mathbf{0}}$ and $Z_{\mathbf{0},\mathbf{n}_2}$.
Starting point is again the grand canonical Hilbert series.
We assume that the magnetic background charge $\mathbf{M}$ is maximally shifted such that all $\mathbf{M}_{\alpha, a} \geq 0$ and at least one $\mathbf{M}_{\alpha,a} = 0$.
As for $L = K = 1$, the grand canonical Hilbert series when expanded in $q_1$ first is
\begin{equation}
  H^{[\mathbf{M}]}(\Lambda) = \sum_{N = 0}^{\infty} \Lambda^N H_N^{[\mathbf{M}]}
  = H_{\infty}
  \sum_{\mathbf{n} \in [\mathbf{M}]_+} \frac{1}{1 - \Lambda q_2^{|\mathbf{n}|}}
  Z_{\mathbf{0},\mathbf{n}}.
\end{equation}
The coefficient $Z_{\mathbf{0},\mathbf{m}}$ for
$\mathbf{m} \in [\mathbf{M}]_+$ is extracted by analytically continuing
to $\Lambda > 1$ and extracting the residue at $q_2^{-|\mathbf{m}|}$.
We shift the residue to $\Lambda = 1$ by redefining
$\Lambda \to \Lambda q_2^{-|\mathbf{m}|}$.
Since $H_N^{[\mathbf{M}]} = H_N^{[\mathbf{m}]}$, we may write
\begin{equation}
  q_2^{-N |\mathbf{m}|} H_N^{[\mathbf{M}]} =
  q_2^{-N |\mathbf{m}|} H_N^{[\mathbf{m}]} = \bar{H}_N^{\mathbf{m}}
\end{equation}
such that
\begin{equation}
  H^{[\mathbf{M}]}(\Lambda q_2^{-|\mathbf{m}|}) =
  \sum_{N = 0}^{\infty} \Lambda^N \bar{H}_N^{\mathbf{m}}
  \equiv \bar{H}^{\mathbf{m}}(\Lambda).
\end{equation}
Similarly for expanding in $q_2$,
\begin{equation}
  H^{[\mathbf{M}]}(\Lambda q_1^{-|\mathbf{m}|}) =
  \sum_{N=0}^{\infty} \Lambda^N
  q_1^{-N |\mathbf{m}|} H_N^{[\mathbf{M}]} =
  \sum_{N=0}^{\infty} \Lambda^N \bar{H}_N^{-\mathbf{m}}
  \equiv \bar{H}^{-\mathbf{m}}(\Lambda),
\end{equation}
where we used $[\mathbf{M}] = [\mathbf{-m}]$.
In other words,
\begin{align}
  Z_{\mathbf{0},\mathbf{m}} &=
  - \frac{1}{H_{\infty}} \operatorname{Res}_{\Lambda = 1}
  \bar{H}^{\mathbf{m}}(\Lambda) \quad \text{for} \quad
  \mathbf{m} \in [\mathbf{M}]_+, \\
  Z_{\mathbf{m},\mathbf{0}} &=
  - \frac{1}{H_{\infty}} \operatorname{Res}_{\Lambda = 1}
  \bar{H}^{-\mathbf{m}}(\Lambda) \quad \text{for} \quad
  \mathbf{m} \in [-\mathbf{M}]_+.
\end{align}
Next, we will calculate the residue. The upshot is that
\begin{align}
  \label{eq:2509151315}
  Z_{\mathbf{0},\mathbf{m}}(q_1, q_2; \vec{x}) &= \frac{\bar{H}_{\infty}^{\mathbf{m}}(q_1, q_2; \vec{x})}{H_{\infty}(q_1, q_2; \vec{x})}
  = \frac{r(\mathbf{m}^T, q^{-1}, \vec{q}_1^L)}{r(\mathbf{0}, q^{-1}, \vec{q}_1^L)}
  \quad \text{and} \\
  Z_{\mathbf{m},\mathbf{0}}(q_1, q_2; \vec{x}) &= \frac{\bar{H}_{\infty}^{-\mathbf{m}}(q_1, q_2; \vec{x})}{H_{\infty}(q_1, q_2; \vec{x})}
  = \frac{r(\mathbf{m}^T, q^{-1}, \vec{q}_2^L)}{r(\mathbf{0}, q^{-1}, \vec{q}_2^L)},
\end{align}
where the components of $\vec{q}_1^L$ and $\vec{q}_2^L$ are
\begin{equation}
  q_{1, \alpha}^L = q^{L_{\alpha} / 2} x_{\alpha}
  \quad \text{and} \quad
  q_{2, \alpha}^L = q^{L_{\alpha} / 2} x_{\alpha}^{-1}.
\end{equation}
$\mathbf{m}^T$ is obtained by transposing the Young diagrams corresponding to the individual $\mathbf{m}_{\alpha}$,
i.e.~$\mathbf{m}^T = (\mathbf{m}_0^T, \ldots, \mathbf{m}_{K-1}^T)$
(recall that $\mathbf{m}_{\alpha,1} \geq \mathbf{m}_{\alpha,2} \geq \ldots \geq \mathbf{m}_{\alpha,L_{\alpha}} \geq 0$, so $\mathbf{m}_{\alpha}$ can be thought of as a partition of $|\mathbf{m}_{\alpha}|$ into at most $L_{\alpha}$ parts).
The function $r$ is the Hausel generating function introduced in appendix \ref{sec:fermionic_forms_and_bg_flux}.
We were able to verify the simple-sum expansion
\eqref{eq:2509232110} perturbatively in Mathematica using the coefficients
\eqref{eq:2509151315}.
In section \ref{sec:gen_func_poincare}, we relate the Hausel generating functions
to fermionic forms.

We will focus on  the proof for $\bar{H}_N^{\mathbf{m}}$, but the case of
$\bar{H}_N^{-\mathbf{m}}$ works the same.
Starting point is the observation that the rescaled Hilbert series
$\bar{H}_{N}^{\mathbf{m}}(q_1, q_2; \vec{x})$,
\begin{equation}
  \bar{H}_{N}^{\mathbf{m}}(q_1, q_2; \vec{x}) =
  q^{- N \|\mathbf{m}\| / 2}
	\sum_{\mathbf{n}^\alpha \in \mathbb{Z}^{N} / S_{N}}
	q^{\Delta[\mathbf{n}, \mathbf{m}]}
	\prod_{\alpha = 0}^{K-1}
	x_{\alpha}^{|\mathbf{n}^{\alpha}|} P_{\mathbf{n}^{\alpha}} (q),
\end{equation}
still factorises in the $N \to \infty$ limit as for $L = K = 1$.
As before, we split $\mathbf{n}^{\alpha} \in \mathbb{Z}^{N} / S_{N}$ into
$(\pi^{\alpha}, \mathbf{0}, -\nu^{\alpha})$ with $\pi^{\alpha}$ and $\nu^{\alpha}$ partitions
of length $l_1$ and $l_2$, respectively.
Then,
\begin{equation}
	\begin{aligned}
		\prod_{\alpha= 0}^{K-1} x_{\alpha}^{|\mathbf{n}^{\alpha}|} & =
		\prod_{\alpha=0}^{K-1} x_{\alpha}^{|\pi^{\alpha}| - |\nu^{\alpha}|} \quad \text{and} \\
		\prod_{\alpha = 0}^{K-1} P_{\mathbf{n}^{\alpha}}(q)      & =
		\prod_{\alpha=0}^{K-1} P_{\pi^{\alpha}}(q) P_{\nu^{\alpha}}(q)
		\prod_{l=1}^{N - l_1 - l_2} \frac{1}{1 - q^{l}}.
	\end{aligned}
\end{equation}
Defining
\begin{equation}
	h(\lambda, \mu) = \frac{1}{2} C_{\alpha \beta}
  \sum_{a=1}^{l(\lambda^{\alpha})} \sum_{b=1}^{l(\lambda^{\beta})} B(\lambda_a^{\alpha}, \lambda_b^{\beta})
	- \sum_{\alpha = 0}^{K-1} \sum_{a=1}^{l(\lambda^\alpha)} \sum_{b=1}^{l(\mu_{\alpha})} B(\lambda_a^{\alpha}, \mu_{\alpha,b})
\end{equation}
(with an implicit sum over $\alpha$, $\beta$ in the first term)
for partitions $\lambda^{\alpha}, \mu_{\alpha} \in \mathcal{P}$ and with $B(a,b)$ as in eq.~\eqref{eq:B_min_zero},
the monopole charge splits into
\begin{equation}
	\begin{aligned}
		\Delta[\mathbf{n}, \mathbf{m}] & =
		h(\pi, \mathbf{m}^+) + h(\nu, \mathbf{m}^-) - \frac{1}{4}
    C_{\alpha \beta}
		\sum_{a,b = 1}^N (|\mathbf{n}_a^{\alpha}| + |\mathbf{n}_b^{\beta}|)
		+ \frac{1}{2} \sum_{\alpha = 0}^{K-1} \sum_{a = 1}^N \sum_{b=1}^{L_{\alpha}}  (|\mathbf{n}_a^{\alpha}| + |\mathbf{m}_{\alpha, b}|)                                       \\
		                               & = h(\pi, \mathbf{m}^+) + h(\nu, \mathbf{m}^-)
		+ \frac{1}{2} \sum_{\alpha=0}^{K-1} (L_{\alpha} |\pi^{\alpha}| + L_{\alpha} |\nu^{\alpha}| + N \| \mathbf{m}_{\alpha} \|).
	\end{aligned}
\end{equation}
In the first two terms, $\mathbf{m}^{\pm} = (\mathbf{m}_0^{\pm}, \ldots, \mathbf{m}_{K-1}^{\pm})$
are the tuples of partitions obtained by
splitting all $\mathbf{m}_{\alpha}$ into positive, zero and negative parts $(\mathbf{m}_{\alpha}^+, \mathbf{0}_{\alpha}, -\mathbf{m}_{\alpha}^-)$.
The third term vanishes since
\begin{equation}
	\sum_{\alpha=0}^{K-1} C_{\alpha \beta} = 0
\end{equation}
for any column $\beta$ of the generalised Cartan matrix.
Then,
\begin{equation}
	\begin{aligned}
    \bar{H}_{N}^{\mathbf{m}}(q_1, q_2; \vec{x}) =&
		       \sum_{0 \leq l_1^\alpha + l_2^\alpha \leq N}
           \left[\sum_{\substack{\pi^{\alpha} \in \mathcal{P}\\ l(\pi^{\alpha}) = l_1^{\alpha}}}
		q^{h(\pi, \mathbf{m}^+)} \prod_{\alpha=0}^{K-1} (q_{1, \alpha}^L)^{|\pi^{\alpha}|}
		P_{\pi^{\alpha}}(q)
		\right]                                                           \\
    \times & \left[\sum_{\substack{\nu^{\alpha} \in \mathcal{P}\\ l(\nu) = l_2^{\alpha}}}
		q^{h(\nu, \mathbf{m}^-)} \prod_{\alpha=0}^{K-1} (q_{2, \alpha}^L)^{|\nu^{\alpha}|}
		P_{\nu^{\alpha}}(q)
		\right]
		\left[
		\prod_{\alpha=0}^{K-1} \prod_{l=1}^{N^{\alpha} - l_1^{\alpha} - l_2^{\alpha}} \frac{1}{1 - q^{l}}
		\right]                                                           \\
		\equiv & \sum_{0 \leq l_1^{\alpha} + l_2^{\alpha} \leq N}
		r_{\{l_1^{\alpha}\}}(\mathbf{m}^+, q^{-1}, \vec{q}_1^L)
		r_{\{l_2^{\alpha}\}}(\mathbf{m}^-, q^{-1}, \vec{q}_2^L)
		r_{\{N - l_1^{\alpha} - l_2^{\alpha}\}}^{(0)}(q)
	\end{aligned}
\end{equation}
There is an obstruction to factorising the grand canonical rescaled Hilbert series $\bar{H}^{\mathbf{m}}(\Lambda)$ as in the previous case because of the single sum over $N$.
To factorise, it would be necessary to sum over unequal gauge ranks $N^{\alpha}$ at the nodes.
We implement this by first summing over the general case with unequal gauge ranks and then projecting the result back onto the case of $N^{\alpha} = N$.
For that, introduce fugacities $\Lambda_{\alpha}$ corresponding to $N^{\alpha}$.
With $\Lambda = \Lambda_0 \cdots \Lambda_{K-1}$, the grand
canonical Hilbert series can be written as
\begin{equation}
  \bar{H}^{\mathbf{m}}(\Lambda) =
  \sum_{N^0 = \ldots = N^{K-1} = 0}^{\infty} 
  \left(\prod_{\alpha = 0}^{K-1} \Lambda_{\alpha}^{N^{\alpha}}\right)
  \bar{H}_{N^0, \ldots, N^{K-1}}^{\mathbf{m}},
\end{equation}
where
\begin{equation}
  \label{eq:251101095}
  \bar{H}_{N^0, \ldots, N^{K-1}}^{\mathbf{m}} =
   \sum_{0 \leq l_1^{\alpha} + l_2^{\alpha} \leq N^{\alpha}}
		r_{\{l_1^{\alpha}\}}(\mathbf{m}^+, q^{-1}, \vec{q}_1^L)
		r_{\{l_2^{\alpha}\}}(\mathbf{m}^-, q^{-1}, \vec{q}_2^L)
		r_{\{N^{\alpha}- l_1^{\alpha} - l_2^{\alpha}\}}^{(0)}(q)
\end{equation}
with
\begin{equation}
	\label{eq:r_l_general}
	r_{l^0,\ldots,l^{K-1}}(\mathbf{m}, q^{-1}, y) =
  \sum_{\substack{\pi^{\alpha} \in \mathcal{P}\\ l(\pi^{\alpha}) = l^{\alpha}}}
	q^{h(\pi, \mathbf{m})} \prod_{\alpha=0}^{K-1} y_{\alpha}^{|\pi^{\alpha}|} P_{\pi^{\alpha}}(q)
\end{equation}
and
\begin{equation}
  r_{l^0, \ldots, l^{K-1}}^{(0)}(q) = \prod_{\alpha = 0}^{K-1} \prod_{l=1}^{l^\alpha} \frac{1}{1 - q^{l}}.
\end{equation}
Although we will not make use of it, \eqref{eq:251101095} is indeed the (rescaled) Coulomb branch formula for the Hilbert series of the theory with unequal gauge ranks.
The generating function without the restriction of $N^{\alpha} = N$ is
\begin{equation}
  \bar{H}^{\mathbf{m}}(\Lambda_{\alpha}) =
  \bar{H}^{\mathbf{m}}(\Lambda_0, \ldots, \Lambda_{K-1}) \equiv
  \sum_{N^0, \ldots, N^{K-1} = 0}^{\infty} 
  \left(\prod_{\alpha = 0}^{K-1} \Lambda_{\alpha}^{N^{\alpha}} \right) \bar{H}_{N^0, \ldots, N^{K-1}}^{\mathbf{m}}.
\end{equation}
We project from $\bar{H}^{\mathbf{m}}(\Lambda_{\alpha})$ onto $\bar{H}^{\mathbf{m}}(\Lambda)$ in the following way.
Define new fugacities as
\begin{equation}
  \label{eq:2509151258}
    \Lambda_0 = \Lambda \zeta_K / \zeta_1
    \quad \text{and} \quad
    \Lambda_{\alpha} = \zeta_{\alpha} / \zeta_{\alpha+1}, \quad \alpha = 1, \ldots, K-1.
\end{equation}
Since the grand canonical Hilbert series at unequal ranks is defined for $|\Lambda_{\alpha}| < 1$, this is a valid
expansion for
\begin{equation}
  \label{eq:2509150915}
|\zeta_{\alpha}| < |\zeta_{\alpha + 1}| \quad (\alpha = 1, \ldots, K-1)
\quad \text{and} \quad
|\zeta_K| < |\zeta_1 / \Lambda|,
\end{equation}
so in particular also $|\Lambda| < 1$.
The Hilbert series at equal ranks $N^{\alpha} = N$ consists of exactly those
terms independent of $\zeta_{\alpha}$, which are extracted by a contour integral as
\begin{equation}
  \bar{H}^{\mathbf{m}}(\Lambda) = 
  \prod_{\alpha = 0}^{K-1} \oint \frac{d \zeta_{\alpha}}{2 \pi i \zeta_{\alpha}} \bar{H}^{\mathbf{m}}(\Lambda_0, \ldots, \Lambda_{K-1})
\end{equation}
on concentric circles around $\zeta_{\alpha} = 0$, such that the conditions \eqref{eq:2509150915} are fulfilled.
Then,
\begin{equation}
  \operatorname{Res}_{\Lambda = 1} \bar{H}^{\mathbf{m}}(\Lambda) = 
  \left[\oint_{|\Lambda - 1| = \epsilon} \frac{d \Lambda}{2 \pi i} \prod_{\alpha = 0}^{K-1} \oint \frac{d \zeta_{\alpha}}{2 \pi i \zeta_{\alpha}}\right] \bar{H}^{\mathbf{m}}(\Lambda_0, \ldots, \Lambda_{K-1}).
\end{equation}
The grand canonical Hilbert series at unequal ranks factorises as
\begin{equation}
  \begin{aligned}
    \bar{H}^{\mathbf{m}}(\Lambda_{\alpha}) 
                                 &=
    \sum_{N^{\alpha} = 0}^{\infty} \sum_{l_{1}^{\alpha} + l_2^{\alpha} + l_3^{\alpha} = N^{\alpha}}^{} 
                                 \left[\Lambda_{0}^{l_1^{0}} \cdots \Lambda_{K-1}^{l_1^{K-1}} r_{\{l_1^{\alpha}\}}^{(+)}(\mathbf{m}^+)\right]
  \left[\Lambda_{0}^{l_2^{0}} \cdots \Lambda_{K-1}^{l_2^{K-1}} r_{\{l_2^{\alpha}\}}^{(-)}(\mathbf{m}^-)\right] \\
                                 &\qquad \qquad \times
  \left[\Lambda_{0}^{l_3^{0}} \cdots \Lambda_{K-1}^{l_3^{K-1}} r^{(0)}_{\{l_3^{\alpha}\}}\right] \\
                                 &= \left[\sum_{l_1^\alpha = 0}^{\infty} \left(\prod_{\alpha = 0}^{K-1} \Lambda_{\alpha}^{l_1^{\alpha}}\right) r_{\{l_1^{\alpha}\}}^{(+)}(\mathbf{m}^+) \right]
                                  \left[\sum_{l_2^\alpha = 0}^{\infty} \left(\prod_{\alpha = 0}^{K-1} \Lambda_{\alpha}^{l_2^{\alpha}}\right) r_{\{l_2^{\alpha}\}}^{(-)}(\mathbf{m}^-) \right] \\
                                 & \qquad \qquad \times \left[\sum_{l_3^\alpha = 0}^{\infty} \left(\prod_{\alpha = 0}^{K-1} \Lambda_{\alpha}^{l_3^{\alpha}}\right) r_{\{l_3^{\alpha}\}}^{(0)} \right] \\
                                 &\equiv r^{(+)}(\Lambda_{\alpha}; \mathbf{m}^+) r^{(-)}(\Lambda_{\alpha}; \mathbf{m}^-) r^{(0)}(\Lambda_{\alpha}).
  \end{aligned}
\end{equation}
We change the variables in the contour integral from $(\Lambda, \zeta_0, \ldots, \zeta_{K-1})$ to $(\zeta, \Lambda_0, \ldots, \Lambda_{K-1})$ where the integral over $\zeta \equiv (\zeta_0 \cdots \zeta_{K-1})^{1/K}$ decouples. The result is
\begin{equation}
  \label{eq:2509151312}
  \operatorname{Res}_{\Lambda = 1}
  \bar{H}^{\mathbf{m}}(\Lambda) = 
  (-1)^{K-1} \left[\prod_{\alpha = 0}^{K-1} \oint_{|\Lambda_{\alpha} - 1| = \epsilon} \frac{d \Lambda_{\alpha}}{2 \pi i}\right] \bar{H}^{\mathbf{m}}(\Lambda_{\alpha}).
\end{equation}
We show in appendix \ref{sec:fermionic_forms_and_bg_flux} that the factors $r^{(\pm)}$ are regular 
as $\Lambda_{\alpha} \to 1$ and become the
Hausel generating function $r(\mathbf{m}^{\pm T}, q^{-1}, \vec{q}_{1,2})$ in \eqref{eq:hausel_gen_func}.
The third factor $r^{(0)}$ is by the $q$-binomial theorem
\begin{equation}
  r^{(0)}(\Lambda_\alpha) = \prod_{\alpha = 0}^{K-1} \prod_{l=0}^{\infty} \frac{1}{1 - \Lambda_{\alpha} q^{l}}
\end{equation}
and has simple poles when $\Lambda_{\alpha} = 1$. Therefore, the integral in \eqref{eq:2509151312}
yields 
\begin{equation}
  \label{eq:2509151328}
  - \operatorname{Res}_{\Lambda = 1} \bar{H}^{\mathbf{m}}(\Lambda) = 
	r(\mathbf{m}^{T}, q^{-1}, \vec{q}_1^L)
	r(\mathbf{0}, q^{-1}, \vec{q}_2^L)
  \prod_{\alpha=0}^{K-1} \prod_{l=1}^{\infty} \frac{1}{1 - q^{l}}.
\end{equation}
Dividing by
\begin{equation}
  H_{\infty} = - \operatorname{Res}_{\Lambda = 1} \bar{H}^{\mathbf{0}}(\Lambda) =
	r(\mathbf{0}, q^{-1}, \vec{q}_1^L)
	r(\mathbf{0}, q^{-1}, \vec{q}_2^L)
  \prod_{\alpha=0}^{K-1} \prod_{l=1}^{\infty} \frac{1}{1 - q^{l}}
\end{equation}
concludes the proof of eq.~\eqref{eq:2509151315}.

\section{Fermionic Forms and Vertex Algebras}
\label{sec:gen_func_poincare}

In the previous sections we have seen that the coefficients occurring in the simple-sum truncation of the giant graviton expansion take the form
\begin{equation}
  \label{eq:2511011007}
	\frac{r(\nu, q^{-1}, \vec{y})}{r(0, q^{-1}, \vec{y})}.
\end{equation}
The function $r$ is known in a related context: by Hausel's formula \cite{Hausel_2006}, when the $\nu_{\alpha}$ are numbers rather than partitions, the ratio \eqref{eq:2511011007} is the generating function of Poincar\'e polynomials of the moduli spaces of instantons on $\mathbb{C}^2 / \mathbb{Z}_K$.
These instanton moduli spaces are quiver varieties $\mathcal{M}_{n, \nu}$ related to the mirror $\mathcal{T}_N[L, K]$ quiver:
they consist of $K$ gauge nodes arranged in a circle, with gauge ranks $n^{\alpha}$.
The framing of this quiver is given by the $\nu_{\alpha} \in \mathbb{N}_0$.
Hausel's formula states that
\begin{equation}
  \label{eq:2511021507}
	\frac{r(\nu, q^{-1}, \vec{y})}{r(0, q^{-1}, \vec{y})} =
  \sum_{n^{\alpha} = 0}^{\infty} 
  P_q [\mathcal{M}_{n, \nu}]
  q^{-d(n, \nu)} \prod_{\alpha = 0}^{K-1} y_{\alpha}^{n^{\alpha}},
\end{equation}
where
\begin{equation}
  d(n, \nu) = n^{\alpha} \nu_{\alpha} - \frac{1}{2} C_{\alpha \beta}
  n^{\alpha} n^{\beta}
\end{equation}
and $P_q[\mathcal{M}_{n, \nu}]$ is the Poincar\'e polynomial of $\mathcal{M}_{n, \nu}$.
$C_{\alpha \beta}$ is again the generalised Cartan matrix, built from the adjacency matrix of the quiver.
If this is interpreted as a formal expansion in $y_{\alpha} q^{-\nu_{\alpha}}$, the expansion coefficients are polynomials in $q$ with positive integer coefficients.

\paragraph{}
Since the instanton moduli spaces admit the action of a certain quantum algebra, the generating function \eqref{eq:2511021507} can naturally be expanded in $q$-characters of this algebra.
Much of these relations are better understood in the context of non-affine ADE-type quivers.
By the fermionic Lusztig conjecture \cite{lusztig2000}, the generating function \eqref{eq:2511021507} can be expressed in terms of fermionic forms, which Mozgovoy subsequently proved by establishing a combinatorial identity between the function $r$ and the fermionic form $n$ \cite{mozgovoy2007ffqv}.
For partition-valued $\nu_{\alpha}$, it was shown in \cite{Barns-Graham:2018qvx} that \eqref{eq:2511021507} is the fusion product of classical Kirillov-Reshetikhin modules \cite{di2008proof}. When $q \to 1$, this becomes an ordinary tensor product of representations of the algebra.
In this section we observe that much of this goes through for affine quivers as well.


\subsection{Fermionic forms}
\label{sec:fermionic_forms}

The second fermionic form $n(\nu, q, \vec{y})$ for an unoriented quiver with $K$ nodes is defined as
\begin{equation}
	\label{eq:fermionic_form}
	n(\nu, q, \vec{y}) \equiv \sum\limits_{\tau \in \mathcal{P}^K}
  \prod_{k = 1}^{\infty} q^{-(\nu_k, \tau_k)} q^{\frac{1}{2}(\tau_k, \tau_k)}
  \prod_{\alpha = 0}^{K-1}
  y_{\alpha}^{\tau_k^{\alpha}}
	\left[\sum\limits_{i=1}^k (\nu_{\alpha,i} - \tau_{\alpha, i}), \tau_k^{\alpha} - \tau_{k+1}^{\alpha} \right]_q,
\end{equation}
where
\begin{equation}
  [n,m]_q = \prod_{i=1}^{m} \frac{(1-q^{n+i})}{1-q^{i}}
  \quad \text{for} \quad
  n \in \mathbb{Z}, m \in \mathbb{N}_0.
\end{equation}
The variable $\nu$ is a $K$-tuple of partitions, $\nu  = (\nu_0, \ldots, \nu_{K-1})\in \mathcal{P}^K$ and the sum is also over $K$-tuples of partitions $\tau = (\tau^0, \ldots, \tau^{K-1}) \in \mathcal{P}^K$.
We denote both $\nu$ and $\tau$ by double indices  $\nu_{\alpha,k}$ and $\tau_k^{\alpha}$.
Here, $k$ is the $k$-th entry in the partitions $\nu_{\alpha}$ and $\tau^{\alpha}$ (we set this to zero if $k$ is larger than the length of the partition). 
At fixed $k$, $\nu_k \in \mathbb{N}_0^K$ and $\tau_k \in \mathbb{N}_0^K$ are elements of the (positive) weight lattice and root lattice, respectively, of the Kac-Moody algebra described by the quiver.
The inner product between a root and a weight is simply
\begin{equation}
  \label{eq:2511032230}
  (\nu_k, \tau_k) = \nu_{\alpha, k} \tau_k^\alpha.
\end{equation}
A root vector is also a weight vector, with components
\begin{equation}
  \tau_{\alpha,k} = C_{\alpha \beta} \tau_k^{\beta}.
\end{equation}
This is not invertible since in general $C_{\alpha \beta}$ is degenerate.
The inner product between two root vectors is then
\begin{equation}
  (\tau_k, \tau_k) = \tau_{\alpha,k} \tau_k^{\alpha} = C_{\alpha \beta} \tau_k^{\alpha} \tau_k^{\beta}.
\end{equation}
Since the graph is unoriented, the inner product between two roots is symmetric.
From now on, we focus on the circular quiver that is the Dynkin diagram of the affine $\widehat{\mathfrak{su}}(K)$ Lie algebra. Then,
\begin{equation}
  \label{eq:2511032231}
  C_{\alpha \beta} = 2 \delta_{\alpha \beta} - \delta_{\alpha, \beta + 1} - \delta_{\alpha, \beta - 1}
\end{equation}
as in eq.~\eqref{eq:2511031252}.

\paragraph{}
Mozgovoy showed for $\nu \in \mathbb{N}_0^K$ and an underlying non-affine ADE-type quiver that the Hausel generating function $r(\nu, q^{-1}, \vec{y})$
is related to the second fermionic form by the identity \cite{mozgovoy2007ffqv}
\begin{equation}
	\label{eq:f_form-identity}
	n(\nu, q, \vec{y}) = r(\nu, q^{-1}, \vec{y}) \, r(0, q, \vec{y})
\end{equation}
and in \cite{Barns-Graham:2018qvx} it was shown that the identity \eqref{eq:f_form-identity} holds generalised to partitions $\nu \in \mathcal{P}^K$.
We have done extensive Mathematica checks to verify that \eqref{eq:f_form-identity} transfers to affine quivers like $\widehat{\mathfrak{su}}(K)$ as well. 
We expect that the proof is a simple adaptation of \cite{mozgovoy2007ffqv, Barns-Graham:2018qvx}.
Therefore, the ratio \eqref{eq:2511011007} is equal to
\begin{equation}
	\label{eq:f_form-identity-ratio}
	\frac{r(\nu, q^{-1}, \vec{y})}{r(0, q^{-1}, \vec{y})} =
	\frac{n(\nu, q, \vec{y})}{n(0, q, \vec{y})}.
\end{equation}
In appendix \ref{sec:fermionic_forms-adhm_quiver} we
derive an analytic formula for the ratio \eqref{eq:f_form-identity-ratio} of fermionic forms in the case of the ADHM quiver with $L = K = 1$ and show that this reproduces the result \eqref{eq:Z_n_0}.

\paragraph{}
The numerator can be written in a simple form:
Hua's formula \cite{hua2000} expresses $r(\nu, q, \vec{y})$ when $\nu = 0$ as the plethystic exponential
\begin{equation}
  r(0,q,\vec{y}) = \operatorname{PE}\left[\frac{a(q,\vec{y})}{q-1}\right].
\end{equation}
The function $a(q, \vec{y})$ is the generating function of the number of ``absolutely indecomposable representations'' of $\widehat{\mathfrak{su}}(K)$ of dimension $\lambda$,
\begin{equation}
  a(q, \vec{y}) = \sum_{\lambda \in \mathbb{N}_0^K} a_{\lambda}(q) y^{\lambda}.
\end{equation}
Kac showed \cite{kac2006} that the coefficients $a_{\lambda}(q)$ are non-zero if and only if $\lambda$ is a positive root of $\widehat{\mathfrak{su}}(K)$.
Furthermore, if $\lambda$ is a real positive root, $a_{\lambda}(q) = 1$ and for imaginary roots $a_{\lambda}(q) = q + K-1$.
The fugacities for the simple roots of $\mathfrak{su}(K)$ are the $K-1$ fugacities $y_{\alpha}$ for
$\alpha = 1, \ldots, K-1$. We change to the more conventional basis by introducing $K$ fugacities $z_{\alpha}$ subject to the constraint $\prod_{\alpha = 1}^{K} z_{\alpha} = 1$, which are related to the $\vec{y}$ by
\begin{equation}
  y_{\alpha} = z_{\alpha} / z_{\alpha+1} \quad \text{for} \quad
  \alpha = 1, \ldots, K-1.
\end{equation}
The fugacity for the imaginary root of $\widehat{\mathfrak{su}}(K)$ is
\begin{equation}
  y = \prod_{\alpha = 0}^{K-1} y_{\alpha}.
\end{equation}
This means that
\begin{equation}
  a(q, \vec{y}) = \sum_{\alpha < \beta}^{} \frac{z_{\alpha}}{z_{\beta}} +
  \frac{y}{1-y} \sum_{\alpha \not= \beta}^{K} \frac{z_{\alpha}}{z_{\beta}} +
  \frac{y}{1-y} (q + K - 1)
\end{equation}
The fermionic form $n(0,q,\vec{y})$ at $\nu = 0$ is a product of the two $r$ functions $r(0, q^{\pm1}, \vec{y})$.
Their dependence on $q$ cancels each other, leaving $n(0,q, \vec{y})$ independent of $q$:
\begin{equation}
  n(0, q, \vec{y}) =
  \operatorname{PE}\left[\frac{a(q^{-1}, \vec{y})}{q^{-1}-1} 
  +
  \frac{a(q, \vec{y})}{q-1} \right]
  = \operatorname{PE} \left[\frac{y}{1-y} \right]
  \operatorname{PE}\left[
    -a(0, \vec{y})
    \right].
\end{equation}
the first factor is the Pochhammer symbol
\begin{equation}
  \operatorname{PE} \left[\frac{y}{1-y}\right] =
  \prod_{k=1}^{\infty} \frac{1}{1 - y^k} = \frac{1}{(y; y)_{\infty}}
\end{equation}
and the function $a(0, \vec{y})$ is just a sum over all positive roots with their respective multiplicities.
The plethystic exponential gives the Weyl denominator of $\widehat{\mathfrak{su}}(K)$,
\begin{equation}
  \operatorname{PE}[-a(0,\vec{y})] =
  \operatorname{PE}\left[- \sum_{\alpha > 0}^{} y^{\alpha}\right] =
  \prod_{\alpha > 0}^{} (1 - y^{\alpha}) \equiv
  \Delta_{\widehat{\mathfrak{su}}(K)}(\vec{y}).
\end{equation}
Put together,
\begin{equation}
  \label{eq:2511031805}
  n(0, q, \vec{y}) =
  \frac{\Delta_{\widehat{\mathfrak{su}}(K)}(\vec{y})}{(y; y)_{\infty}}
\end{equation}
such that we arive at the formula
\begin{equation}
  \label{eq:2511041200}
  \frac{r(\nu, q^{-1}, \vec{y})}{r(0, q^{-1}, \vec{y})} =
  (y; y)_{\infty} \, \frac{n(\nu, q, \vec{y})}{\Delta_{\widehat{\mathfrak{su}}(K)}(\vec{y})}
\end{equation}
for the giant graviton coefficients.

\subsection{Euler character limit}

We refer to the limit $q \to 1$ as the \emph{unrefined limit}.
Whenever the ratio (\ref{eq:2511011007}) has an interpretation in terms of Poincar\'e polynomials as in eq.~(\ref{eq:2511021507}), this limit reduces the Poincar\'e polynomial to the Euler characteristic of the quiver variety.
The Euler characteristic has a combinatorial interpretation as counting the number of fixed points under a certain torus action.
For instanton moduli spaces, these are labelled by tuples of Young diagrams of fixed weights.
Numerically to high orders we find that a similar interpretation in the unrefined limit holds for general partition-vectors $\nu \in \mathcal{P}^K$.
Only, here the counting is over tuples of height-restricted plane partitions, which contains ordinary partitions as the special case when all height-restrictions are one.

\paragraph{}
When $K = 1$ and for $\nu = \mathbf{m}^T \in \mathcal{P}$, this is straightforward: we show in (\ref{eq:2511032139}) that the ratio (\ref{eq:2511041200}) is
\begin{equation}
  (y; y)_{\infty} \, n(\nu, 1, y) = 
  \prod_{k=1}^{\infty} \frac{1}{(1 - y^k)^{\tilde{\nu}_k}},
  \quad \text{where} \quad
  \tilde{\nu}_k = \sum_{l=1}^{k} \nu_l.
\end{equation}
Given that $\nu_l = \sum_{m = l}^{\infty} \mu_m(\mathbf{m})$, this can be written as
\begin{equation}
  \prod_{k=1}^{\infty} \frac{1}{(1 - y^k)^{\sum_{l=1}^{k} \sum_{m = l}^{\infty} \mu_m(\mathbf{m})}}
  =
  \prod_{k=1}^{\infty} \prod_{m=1}^{\infty} \frac{1}{(1 - y^k)^{\mu_m(\mathbf{m}) \min(m,k)}},
\end{equation}
which is the product of $L = l(\mathbf{m})$ generating functions of plane partitions with maximal height $m \in \mathbf{m}$,
\begin{equation}
  (y; y)_{\infty} \, n(\mathbf{m}^T, 1, y) = 
  \prod_{m \in \mathbf{m}}^{}  
  \prod_{k=1}^{\infty} \frac{1}{(1 - y^k)^{\min(m, k)}}.
\end{equation}
When $K > 1$, the fermionic forms are related to instanton moduli spaces of instantons on $\mathbb{C}^2 / \mathbb{Z}_K$, the effect of which is to introduce a $K$-colouring for the $L$ plane partitions.
We find that the colouring is independent of the height and, for a box $(i, j, k) \in Y$ with maximal height $k \leq h(Y)$, is given by
$\sigma(Y) + i - j \mod K$, where $\sigma(Y)$ is the starting colour of the $(1,1,1)$ box in $Y$.
Denote by $\text{PP}$ the set of all plane partitions and by $|Y|$ for $Y \in \text{PP}$ the weight of the plane partition.
We claim that
\begin{equation}
  (y; y)_{\infty} \, \frac{n(\mathbf{m}^T, 1, \vec{y})}{\Delta_{\widehat{\mathfrak{su}}(K)}(\vec{y})}
  =
  \sum_{\substack{Y_{\alpha,a} \in \text{PP} \\ h(Y_{\alpha,a}) \leq \mathbf{m}_{\alpha,a} \\ \sigma(Y_{\alpha,a}) = \alpha}}^{} 
  \prod_{\alpha = 0}^{K-1} \prod_{a = 1}^{L_{\alpha}} y_{\alpha}^{|Y_{\alpha,a}|}.
\end{equation}
This obviously factorises into
\begin{equation}
  (y; y)_{\infty} \, \frac{n(\mathbf{m}^T, 1, \vec{y})}{\Delta_{\widehat{\mathfrak{su}}(K)}(\vec{y})}
  =
  \prod_{\alpha = 0}^{K-1} \prod_{a = 1}^{L_{\alpha}} \left[(y;y)_{\infty} \frac{n(\mathbf{m}_{\alpha,a}^T, 1, \vec{y})}{\Delta_{\widehat{\mathfrak{su}}(K)}(\vec{y})} \right],
\end{equation}
which can readily be checked.
Furthermore, there is numerical evidence that when $q \not= 1$, this is a $q$-counting of plane partitions,
\begin{equation}
  (y; y)_{\infty} \, \frac{n(\mathbf{m}^T, q, \vec{y})}{\Delta_{\widehat{\mathfrak{su}}(K)}(\vec{y})}
  =
  \sum_{\substack{Y_{\alpha,a} \in \text{PP} \\ h(Y_{\alpha,a}) \leq \mathbf{m}_{\alpha,a} \\ \sigma(Y_{\alpha,a}) = \alpha}}^{} 
  q^{\#(Y)}
  \prod_{\alpha = 0}^{K-1} \prod_{a = 1}^{L_{\alpha}} y_{\alpha}^{|Y_{\alpha,a}|}.
\end{equation}
As mentioned in a footnote in the introduction, the power $\#(Y)$ is not strictly positive and only so if the expansion variables $y_{\alpha}$ are shifted to $y_{\alpha} q^{-L_{\alpha}}$. This is consistent with eq.~(\ref{eq:2511021507}), where the coefficients are not directly polynomials in $q$ but come with an overall power of $q^{-d(n, \nu)}$ that is negative for a finite number of terms.

\subsection{Burge conditions}

A plane partition of maximal height $m$ is equivalent to an $m$-tuple of ordinary partitions which are contained in one another, $Y_1 \supseteq Y_2 \supseteq \ldots \supseteq Y_m$.
This condition on the $m$ partitions can be seen as a special case of the \emph{Burge conditions} \cite{BURGE1993210} on a set of ordinary partitions, which arise naturally in the AGT correspondence for minimal models \cite{alday2010, Manabe:2020etw}.
In this context, the AGT correspondence relates 4d $\mathcal{N} = 2$ $U(m)$ super Yang-Mills on $\mathbb{C}^2 / \mathbb{Z}_K$ with Omega-deformation to a two-dimensional CFT with symmetry algebra
\begin{equation}
	\mathcal{V}(m,K; p) = \mathcal{H} \oplus \widehat{\mathfrak{su}}(K)_m
	\oplus \underbrace{\frac{\widehat{\mathfrak{su}}(m)_K \oplus \widehat{\mathfrak{su}}(m)_{p-m}}
		{\widehat{\mathfrak{su}}(m)_{K+p-m}}}_{\equiv W_{m,K}^{\text{para}}},
\end{equation}
where $\mathcal{H}$ is the Heisenberg algebra and $W_{m,K}^{\text{para}}$ is the $K$-th parafermion $W_m$-algebra.
The parameter $p$ is related to the deformation parameters,
\begin{equation}
  p \, \epsilon_1 + (p+K) \, \epsilon_2 = 0
  \quad \text{such that} \quad
  \frac{\epsilon_1}{\epsilon_2} = - 1 - \frac{K}{p}.
\end{equation}
The Yang-Mills instanton partition function is a sum over fixed points in the instanton moduli spaces corresponding to an $m$-tuple of partitions.
It is also the character of a certain representation of the algebra $\mathcal{V}(m, K; p)$.

\paragraph{}
For $p \geq m$ integer, the instanton partition function exhibits unphysical poles which can be removed by restricting the sum over fixed points to Young diagrams satisfying the Burge conditions
\begin{equation}
	\begin{aligned}
		(Y_{i})_j                   & \geq (Y_{i + 1})_{j + r_{i} - 1} - s_{i} + 1
		                            &                                              & \text{and}                                 \\
		\sigma_{i} - \sigma_{i + 1} & \equiv s_{i} - r_{i}
		\mod K
		                            &                                              & \text{for} \quad i \in \{0, \ldots, m-1\}, \\
	\end{aligned}
\end{equation}
where $Y_0 \equiv Y_m$ and $\sigma_{i} = \sigma(Y_i) \in \mathbb{Z}_K$ is the colour of $Y_{i}$.
The possible Burge conditions are parametrised by strictly positive integers $\mathfrak{r} = [r_0, \ldots, r_{m-1}] \in \mathbb{N}^m$ and $\mathfrak{s} = [s_0, \ldots, s_{m-1}] \in \mathbb{N}^m$, subject to the constraints
\begin{equation}
	\sum_{i = 0}^{m-1} r_i = p
  \quad \text{and} \quad
  \sum_{i=0}^{m-1} s_i = K+p.
\end{equation}
In the CFT, the ``forbidden'' Young diagrams correspond to certain null states, and restricting to Burge conditions means restricting to the $(p, p+K)$-minimal model of $W_{m,K}^{\text{para}}$.

\paragraph{}
The case most relevant to our discussion is the Burge condition
\begin{equation}
  \mathfrak{r} = [p + 1 - m, 1, \ldots, 1]
  \quad \text{and} \quad
  \mathfrak{s} = [K + p + 1 - m, 1, \ldots, 1].
\end{equation}
The constraints then become
\begin{align}
  (Y_i)_j &\geq (Y_{i+1})_j & & \text{for } i = 1, \ldots, m-1, \\
  (Y_m)_j &\geq (Y_1)_{j + p - m} - K - p + m & & \text{and} \label{eq:2511041314}\\
  \sigma_i &= \sigma_j & & \text{for } i, j = 1, \ldots, m.
\end{align}
In the limit $p \to \infty$, (\ref{eq:2511041314}) becomes trivial and the Burge conditions are exactly the $\mathbb{Z}_K$-coloured plane partitions of maximal height $m$.
By choosing suitable $\mathfrak{r}$ and $\mathfrak{s}$, one may also embed $L$-tuples of plane partitions into the space of $|\mathbf{m}| = \sum_{\alpha = 0}^{K-1} \sum_{a=1}^{L_{\alpha}} \mathbf{m}_{\alpha,a}$ partitions in a similar way.

\subsection{Affine Lie algebra characters}

Based on the discussion of Burge conditions, it is clear that the giant graviton coefficients, if their interpretation as counting height-restricted plane partitions is correct, are related to characters of the algebra $\mathcal{V}(m, K; p)$ in the $p \to \infty$ limit.
Since the contributions of individual plane partitions factorises in the unrefined limit $q \to 1$, we will focus on a single such plane partition with maximal height $m$.
Plane partitions at different nodes are related by cyclically permuting the fugacities $y_{\alpha} \to y_{\alpha + i}$.

\paragraph{Case $m = 1$ and $K$ arbitrary:}
If $m = 1$, the algebra becomes
\begin{equation}
  \mathcal{V} = \mathcal{H} \oplus \widehat{\mathfrak{su}}(K)_1.
\end{equation}
It was already shown in \cite{Fujii:2005dk} that the generating function of Euler characteristics for $m = (1, 0, \ldots, 0) \in \mathbb{N}_0^K$ is
\begin{equation}
  (y; y)_{\infty} \, \frac{n(m^T, 1, \vec{y})}{\Delta_{\widehat{\mathfrak{su}}(K)}(\vec{y})} =
  \frac{1}{(y; y)_{\infty}} \chi_{\text{vac}}^{\widehat{\mathfrak{su}}(K)_1}(\vec{y}),
\end{equation}
where the character of the vacuum representation of $\widehat{\mathfrak{su}}(K)$ at level $1$ is
\begin{equation}
  \chi_{\text{vac}}^{\widehat{\mathfrak{su}}(K)_1}(\vec{y}) =
	\frac{1}{(y; y)_{\infty}^{K-1}}
	\sum_{\substack{m_0, \ldots, m_{K-1} \in \mathbb{Z},\\
			m_0 + \ldots + m_{K-1} = 0}}
	y^{\sum_{i=0}^{K-1} m_i^2 / 2}
	\prod_{\alpha = 1}^{K-1} y_{\alpha}^{\sum_{i=1}^{\alpha} m_i}.
\end{equation}
As before, $y = \prod_{\alpha = 0}^{K-1} y_{\alpha}$ is the fugacity of the imaginary root of $\widehat{\mathfrak{su}}(K)$ and $y_{\alpha}$ for $\alpha = 1, \ldots, K-1$ are the fugacities for the non-affine $\mathfrak{su}(K)$.
This matches the findings of \cite{Hayashi:2024aaf}.

\paragraph{}
In fact, \cite{Fujii:2005dk} also gave an expression for the generating function Poincar\'e polynomials.
However, this is difficult to compare because their generating function does not contain the $q^{-d(n, \nu)}$ as in Hausel's formula eq.~(\ref{eq:2511021507}).
By the outer $\mathbb{Z}_K$ automorphism of $\widehat{\mathfrak{su}}(K)$, the
other rank one cases $m = (0, \ldots, 1, \ldots, 0) \in \mathbb{N}_0^K$ are
related to the above by cyclically permuting the variables,
\begin{equation}
	y_{\alpha} \to y_{\alpha + i}
\end{equation}
where $y_{\alpha} \sim y_{\alpha + K}$.

\paragraph{Case $K = 1$ and $m$ arbitrary:}
When $K = 1$, the algebra $\mathcal{V}(m, 1; p)$ simplifies to
\begin{equation}
  \mathcal{H} \oplus W_m,
\end{equation}
where $W_m = W_{m,1}^{\text{para}}$ is the (ordinary) $W_m$-algebra.
If further $m = 1$, then $\mathcal{V}(1,1) = \mathcal{H}$ and the character is indeed
\begin{equation}
	\chi_H(y) = \prod_{k=1}^{\infty} \frac{1}{1 - y^k} = \frac{1}{(y; y)_{\infty}}.
\end{equation}
For $m=2$, $W_2 = \operatorname{Vir}$ is the Virasoro algebra. In the $p \to \infty$ limit, the $(p, p+1)$-minimal models have
(normalised) characters
\begin{equation}
	\chi_{r,s}(y) = \frac{1 - y^{rs}}{(y; y)_{\infty}}
\end{equation}
where $r$ and $s$ are related to the Burge conditions by $\mathfrak{r} = [p-r,r]$ and $\mathfrak{s} = [p+1-s,s]$. For $r = s = 1$, this
becomes
\begin{equation}
  (y; y)_{\infty} \, n(m^T, 1, y) =
	\prod_{k=1}^{\infty} \prod_{i=1}^{2} \frac{1}{1 - y^{k+i-1}} =
	\frac{1 - y}{(y; y)_{\infty}^2} =
	\chi_H(y) \chi_{1,1}(y).
\end{equation}
Similarly for $m = 3$, $\mathfrak{r} = [p-2, 1, 1]$ and $\mathfrak{s} = [p-1, 1, 1]$, in the limit $p \to \infty$
\begin{equation}
  (y; y)_{\infty} \, n(m^T, 1, y) =
	 \chi_H(y) \chi_{\mathfrak{r}, \mathfrak{s}}^{W_3}(y)
\end{equation}
with the $W_3$-algebra minimal model character
\begin{equation}
	\chi_{\mathfrak{r}, \mathfrak{s}}^{W_3}(y) = \frac{(1-y)^2(1-y^2)}{(y; y)_{\infty}^2}.
\end{equation}
More generally, in appendix \ref{sec:fermionic_forms-adhm_quiver} we show that
\begin{equation}
  \label{eq:adhm-w_alg-vac_rep}
  (y; y)_{\infty} \, n(m^T, 1, y) =
  \frac{1}{(y;y)_{\infty}} \prod_{l=2}^m \frac{1}{(y^l; y)_\infty}
\end{equation}
and the latter factor is indeed the vacuum character of the $W_m$ algebra.

\paragraph{Case $m$ and $K$ arbitrary: }
In the most general case, the giant graviton contribution becomes the vacuum character of the full algebra,
\begin{equation}
  (y; y)_{\infty} \, \frac{n(m^T, 1, \vec{y})}{\Delta_{\widehat{\mathfrak{su}}(K)}(\vec{y})} =
  \lim_{p \to \infty}
  \chi_{\text{vac}}^{\mathcal{V}(m, K; p)}(\vec{y}) \equiv \chi_{\text{vac}}^{\mathcal{V}(m,K)}(\vec{y}).
\end{equation}
This can be decomposed into characters of $W_{m,K}^{\text{para}}$ and $\widehat{\mathfrak{su}}(K)_m$ as \cite{Manabe:2020etw}
\begin{equation}
  \label{eq:a_vac_char}
  \chi_{\text{vac}}^{\mathcal{V}(m,K)}(\vec{y}) =
  \frac{1}{(y; y)_{\infty}} \sum_{\substack{\mathfrak{l} \in \mathbb{N}_0^m,\\ \mathfrak{l}_0 + \ldots + \mathfrak{l}_{m-1} = K,\\
  f(\mathfrak{l}) = 0}} C_{\mathfrak{l}}^{\mathfrak{r}, \mathfrak{s}}(y)
  \chi_{\mathfrak{l}^T}^{\widehat{\mathfrak{su}}(K)_m}(\vec{y}).
\end{equation}
We will explain the ingredients in this formula in detail in the following.
The functions $C_{\mathfrak{l}}^{\mathfrak{r}, \mathfrak{s}}(y)$ are characters of the parafermion $W$-algebra. By the coset construction
\begin{equation}
  W_{m,K}^{\text{para}} = 
\frac{\widehat{\mathfrak{su}}(m)_K \oplus \widehat{\mathfrak{su}}(m)_{p-m}}
		{\widehat{\mathfrak{su}}(m)_{K+p-m}},
\end{equation}
they are the branching functions of
\begin{equation}
  \chi_{\mathfrak{l}}^{\widehat{\mathfrak{su}}(m)_K} \chi_{\mathfrak{r}-\mathfrak{1}}^{\widehat{\mathfrak{su}}(m)_{p-m}} \sim
  \sum_{\mathfrak{s}}^{} C^{\mathfrak{r}, \mathfrak{s}}_{\mathfrak{l}} \, \chi_{\mathfrak{s} - \mathfrak{1}}^{\widehat{\mathfrak{su}}(m)_{K+p-m}},
\end{equation}
where $\mathfrak{1} = [1, \ldots, 1] \in \mathbb{N}_0^m$ is the Weyl vector of $\widehat{\mathfrak{su}}(m)$.
Similarly, the functions $\chi_{\mathfrak{l}^T}^{\widehat{\mathfrak{su}}(K)_m}(\vec{y})$ are characters of $\widehat{\mathfrak{su}}(K)$ at level $m$ with highest weight $\mathfrak{l}^T$.
They are normalised to begin with grade $y^{h_{\mathfrak{l}^T}}$,
\begin{equation}
  \chi_{\mathfrak{l}^T}^{\widehat{\mathfrak{su}}(K)_m}(\vec{y}) =
  y^{h_{\mathfrak{l}^T}} \mathcal{O}(y^0) 
  \quad \text{where} \quad
  h_{\mathfrak{l}^T} = \frac{(\mathfrak{l}^T, \mathfrak{l}^T + 2 \rho)}{2(m+K)}
\end{equation}
and $\rho = [1, \ldots, 1] \in \mathbb{N}_0^K$ is the Weyl vector of $\widehat{\mathfrak{su}}(K)$.

\paragraph{}
To a highest weight $\mathfrak{l}$ of $\widehat{\mathfrak{su}}(m)_K$, one can associate a highest weight $\mathfrak{l}^T$ of $\widehat{\mathfrak{su}}(K)_m$ as following.
First, level-rank duality identifies equivalence classes of representations of $\widehat{\mathfrak{su}}(m)_K$ with equivalence classes of representations of $\widehat{\mathfrak{su}}(K)_m$.
The equivalence classes $[\mathfrak{l}]$ and $[\mathfrak{l}^T]$ are related by simply transposing the associated Young diagram.
Since $[\mathfrak{l}]$ and $[\mathfrak{l}^T]$ contain a different number of elements, there is no bijection between individual representations.

Importantly, $\mathfrak{l}^T$ is not literally the transpose of the associated Young diagram.
The algorithm to obtain $\mathfrak{l}^T$ is as follows.
First, the weight of the Young diagram corresponding to $\mathfrak{l}$ can be written as
\begin{equation}
  \sum_{\alpha = 1}^{m-1} \alpha \mathfrak{l}_{\alpha} \equiv f + c \, m
  \quad \text{for} \quad c \in \mathbb{Z}
  \quad \text{and} \quad
  0 \leq f < \max \{m, K\}.
\end{equation}
If $K$ if sufficiently large, $f$ is not unique.
But once $(f, c)$ is chosen, we map $\mathfrak{l}$ to the highest weight $\mathfrak{l}^{T, f} \equiv a^{-c} \, T(\mathfrak{l})$ of $\widehat{\mathfrak{su}}(K)_m$, where $T(\mathfrak{l})$ is the actual transpose of the Young diagram,\footnote{To be precise, columns of length $K$ in the transposed Young diagram have to be removed.} but the outer automorphism group $\mathbb{Z}_K$ of $\widehat{\mathfrak{su}}(K)$ acts as the cyclic permutation
\begin{equation}
  a : [\lambda_0, \ldots, \lambda_{K-1}] \to [\lambda_{K-1}, \lambda_0, \ldots, \lambda_{K-2}].
\end{equation}
The terms in the branching formula (\ref{eq:a_vac_char}) are restricted to weights $\mathfrak{l}$ that allow the choice $f(\mathfrak{l}) = 0$, which is essentially determined by the representation of $\mathcal{V}$ being the vacuum representation. The highest weight $\mathfrak{l}^T$ is defined as $\mathfrak{l}^T \equiv \mathfrak{l}^{T,0}$.

\subsection{Higgs branch giant gravitons with wrapping number $m = 2$}

In the following subsection we work through the simplest non-trivial example: the Higgs branch giant graviton coefficients of wrapping number $m = 2$ in the ADHM quiver with $K$ flavours.
The branching of the vacuum character of $\mathcal{V}(m,K)$ follows example 3.3 in \cite{Manabe:2020etw}.
The Burge conditions correspond to the vectors $\mathfrak{r} = [p-1,1] \in \mathbb{N}^2$ and $\mathfrak{s} = [p + K - 1, 1] \in \mathbb{N}^2$.
The branching weights $\mathfrak{l}$ are $\widehat{\mathfrak{su}}(2)_K$ weights $\mathfrak{l} = [K - l, l] \in \mathbb{N}_0^2$.
The parafermion $W_{2,K}^{\text{para}}$ $(p, p+K)$-minimal model characters are given by
\begin{equation}
  \begin{aligned}
  C_{\mathfrak{l}}^{\mathfrak{r}, \mathfrak{s}}(y) =
  \lim_{p \to \infty}
  y^{-B_{1,1}} &\sum_{\substack{n=0\\n \equiv l \mod 2}}^{K}
  \hat{c}_{[K-n,n]}^{\mathfrak{l}}(y) \\
              \times&\left(
    \sum_{\substack{k \in \mathbb{Z}\\pk \equiv n/2 \mod 2}}
    y^{B_{2pk+1,1}} -
    \sum_{\substack{k \in \mathbb{Z}\\pk -1 \equiv n/2 \mod 2}}
    y^{B_{2pk+1,-1}}
  \right),
  \end{aligned}
\end{equation}
where
\begin{equation}
B_{r,s} = \frac{((p+K)r-ps)^2}{4 K p (p+K)}.
\end{equation}
The functions $\hat{c}^{\mathfrak{l}}_{\mathfrak{n}}(y)$ are string functions of maximal weight $\mathfrak{n}$ in the $\widehat{\mathfrak{su}}(2)_K$ representation with dominant weight $\mathfrak{l}$.
Relative to the ordinary string functions $\sigma_{\mathfrak{n}}^{\mathfrak{l}}(y)$, they are
normalised as
\begin{equation}
  \hat{c}_{\mathfrak{n}}^{\mathfrak{l}}(y) = y^{h_{\mathfrak{l}} - \frac{(\mathfrak{n}, \mathfrak{n})}{2K}} \sigma_{\mathfrak{n}}^{\mathfrak{l}}(y)
  \quad \text{with} \quad
h_{\mathfrak{l}} = \frac{(\mathfrak{l}, \mathfrak{l} + 2 \rho)}{2(m+K)} =
\frac{l(l+3)}{4(2+K)}.
\end{equation}
Concretely, they are given by \cite{distler1990}
\begin{equation}
  \begin{aligned}
  \hat{c}_{[K-n,n]}^{[K-l,l]}(y) = 
  \frac{y^{h_{\mathfrak{l}} - \frac{n^2}{4K}}}{(y;y)_{\infty}^3}
  & \sum_{k_1, k_2 = 0}^{\infty} 
  (-1)^{k_1 + k_2} y^{\frac{1}{2}k_1(k_1+1) + \frac{1}{2}(k_2(k_2+1) + (K+1) k_1 k_2)} \\
                               &
                               \times \left[y^{\frac{1}{2}k_1(l-n) + \frac{1}{2}k_2(l+n)} -
y^{K+1-l + \frac{1}{2}k_1 (2K+2-l+n) + \frac{1}{2}k_2 (2K+2-l-n)}\right].
  \end{aligned}
\end{equation}
In the limit $p \to \infty$, $B_{2pk+1,\pm1}$ diverges and the term
vanishes unless $k = 0$. Hence, the parafermion characters collapse to
\begin{equation}
  C_{\mathfrak{l}}^{\mathfrak{r}, \mathfrak{s}}(y) =
  \sum_{\substack{n=0\\n\equiv l \mod 2}}^K
  \hat{c}^{[K-l,l]}_{[K-n,n]}(y) \times
  \begin{cases}
    1 & n/2 \equiv 0 \mod 2 \\
    -y^{1/K} & n/2 \equiv 1 \mod 2
  \end{cases}.
\end{equation}

\paragraph{K = 2:}
When $K = 2$, the weight $\mathfrak{l}$ can take the values $[2,0]$, $[1,1]$ and $[0,2]$. Only the first and the third allow the choice $f(\mathfrak{l}) = 0$.
Concretely, $[2,0]$ allows $(f, c) = (0, 0)$ and $[0,2]$ has $(f, c) = (0,1)$. The transpose is then
\begin{equation}
[2,0]^T = [2,0]
\quad \text{and} \quad
[0,2]^T = [0,2].
\end{equation}
Therefore,
\begin{equation}
  \chi_{\text{vac}}^{\mathcal{V}(2,2)}(\vec{y}) = 
  \frac{1}{(y;y)_{\infty}} \left(C_{[2,0]}^{\mathfrak{r}, \mathfrak{s}}(y) \chi_{[2,0]}^{\widehat{\mathfrak{su}}(2)_2}(\vec{y})
  +C_{[0,2]}^{\mathfrak{r}, \mathfrak{s}}(y) \chi_{[0,2]}^{\widehat{\mathfrak{su}}(2)_2}(\vec{y}) \right).
\end{equation}
The parafermion characters are
\begin{equation}
  \begin{aligned}
    C_{[2,0]}^{\mathfrak{r}, \mathfrak{s}}(y) &= 
  \hat{c}_{[2,0]}^{[2,0]}(y) - y^{1/2} \hat{c}_{[0,2]}^{[2,0]}(y) \\
                                              &=
                                              1 + y^2 + y^3 + 3y^4 + 3y^5 + 7y^6 + 8y^7 + 15y^8 + 19y^9 + \ldots
                                              \quad \text{and} \\
    C_{[0,2]}^{\mathfrak{r}, \mathfrak{s}}(z) &= 
    \hat{c}_{[2,0]}^{[0,2]}(y) - y^{1/2} \hat{c}_{[0,2]}^{[0,2]}(y) \\
                                              &=
                                              y^{-1/2} (y^2 + y^3 + 2y^4 + 3y^5 + 5y^6 + 7y^7 + 12y^8 + 16y^9 + \ldots).
  \end{aligned}
\end{equation}

\paragraph{K = 3:}
The dominant weights of $\widehat{\mathfrak{su}}(2)_3$ are $[3,0]$, $[2,1]$, $[1,2]$ and $[0,3]$.
Only $[3,0]$ and $[1,2]$ admit $f = 0$.
The transpose weights are $[3,0]^T = [2,0,0]$ and $[1,2]^T = [0,1,1]$ such that
\begin{equation}
  \chi_{\text{vac}}^{\mathcal{V}(2,3)}(\vec{y}) = \frac{1}{(y;y)_{\infty}} \left(
    C_{[3,0]}^{\mathfrak{r},\mathfrak{s}}(y) \chi_{[2,0,0]}^{\widehat{\mathfrak{su}}(3)_2}(\vec{y})
    + C_{[1,2]}^{\mathfrak{r}, \mathfrak{s}}(y) \chi_{[0,1,1]}^{\widehat{\mathfrak{su}}(3)_2}(\vec{y})
  \right).
\end{equation}
The parafermion characters are
\begin{equation}
  \begin{aligned}
    C_{[3,0]}^{\mathfrak{r}, \mathfrak{s}}(y) &= \hat{c}_{[3,0]}^{[3,0]}(y) - y^{1/3} \hat{c}_{[1,2]}^{[3,0]}(y) \\
                                              &=
                                              1 + y^2 + y^3 + 3y^4 + 3y^5 + 8y^6 + 9y^7 + 18y^8 + 24y^9 + \ldots
                                              \quad \text{and} \\
    C_{[1,2]}^{\mathfrak{r}, \mathfrak{s}}(y) &= 
    \hat{c}_{[3,0]}^{[1,2]}(y) - y^{1/3} \hat{c}_{[1,2]}^{[1,2]}(y) \\
                                              &=
                                              y^{-3/5} (y^2 + y^3 + 3y^4 + 4y^5 + 8y^6 + 12y^7 + 21y^8 + 30y^9 + \ldots).
  \end{aligned}
\end{equation}

\acknowledgments

The authors would like to thank Andy Zhao, Sam Crew, Yutaka Matsuo, Canberk Sanli, Kevin Crisafio and Julius Grimminger for helpful discussions.
This work has been partially supported by STFC consolidated grant ST/X000664/1.
Part of this research was conducted while funded by the Theoretical Sciences Visiting Program (TSVP) at the Okinawa Institute of Science and Technology (OIST).

\appendix

\section{Twisted limit of 6d superconformal index}
\label{sec:6d_sc_index}

The 6d $(2,0)$ $U(n)$ superconformal index is given by
\begin{equation}
  Z_{S^5 \times S^1}^{U(n)}(\tilde{y}_i, \tilde{q}_I) = \operatorname{Tr} \left[
    (-1)^F e^{- \sum_{i=1}^{3} \tilde{\omega}_i J_i} e^{- \sum_{I=1}^{2} \tilde{\Delta}_I Q_I}
		\right],
\end{equation}
where we introduce chemical potentials for the variables in eq.~\eqref{eq:6d_sc_index} by
\begin{equation}
  \tilde{y}_i = e^{- \tilde{\omega}_i} \quad \text{and} \quad
  \tilde{q}_I = e^{- \tilde{\Delta}_I}.
\end{equation}
The localisation result \cite{Kim:2013nva} expresses the 6d superconformal
index instead in terms of fugacities
$\beta$, $m$, $a_i$, $i = 1,2,3$ by
\begin{equation}
	Z_{S^5 \times S^1}^{U(n)}(\beta, m, a_i) = \operatorname{Tr} \left[
		(-1)^F e^{-\beta \left(E - \frac{Q_1 +Q_2}{2}\right)} e^{-\beta \sum_{i=0}^3 a_i J_i} e^{-\beta m(Q_1 - Q_2)}
		\right],
\end{equation}
subject to the BPS relation and fugacity constraint
\begin{align}
  E = 2Q_1 + 2 Q_2 + J_1 + J_2 + J_3 \quad \text{and}\\
  \beta(a_1 + a_2 + a_3) = 0 \quad \mod 4 \pi i \mathbb{Z}.
\end{align}
These two choices of variables are identified as
\begin{equation}
	\begin{aligned}
		\tilde{\omega}_i & = \beta(1 + a_i),                                       \\
		\tilde{\Delta}_1 & = \beta \left( \frac{3}{2} + m \right) \quad \text{and} \\
		\tilde{\Delta}_2 & = \beta \left( \frac{3}{2} - m \right).
	\end{aligned}
\end{equation}
The result from reduction to $\mathbb{CP}^2 \times S^1$ is
the contour integral expression
\begin{equation}
	Z_{S^5 \times S^1}^{U(n)}(\beta, m, a_i) = \frac{1}{n!} \sum_{s \in \mathbb{Z}^n} \oint
	\left[\frac{d \lambda_i}{2 \pi} \right]^n
	e^{-S_0(\lambda, s, \beta)} \prod_{\alpha=1}^3 Z^{(\alpha)}
\end{equation}
where the classical action is
\begin{equation*}
	S_0(\lambda, s, \beta):=
	\sum\limits_{i=1}^N \left(- \beta \frac{s_i^2}{2} + i s_i \lambda_i
	\right).
\end{equation*}
and 
\begin{equation*}
	Z^{(\alpha)} = Z_{\text{Nekrasov}}(\tau^{(\alpha)}, \epsilon_1^{(\alpha)},
	\epsilon_2^{(\alpha)}, m^{(\alpha)}; i \lambda_i^{(\alpha)}) \equiv
	Z_{\text{pert}}^{(\alpha)} Z_{\text{inst}}^{(\alpha)}
\end{equation*}
is the Nekrasov partition function. The parameters are
\begin{equation}
	\begin{aligned}
		\tau^{(1)}       & = \frac{i}{2 \pi} \beta (1 + a_1), \\
		\epsilon_1^{(1)} & = \beta (a_2 - a_1),                                                                                           \\
		\epsilon_2^{(1)} & = \beta (a_3 - a_1),                                                                                           \\
		m^{(1)}          & = \beta \left(m - \frac{1 + a_1}{2}\right)
		\quad \text{and}                                                                                                                                    \\
		\lambda_i^{(1)}  & = \lambda_i + i s_i \beta a_1
  \end{aligned}
\end{equation}
for $\alpha = 1$. The parameters for $\alpha = 2,3$ are obtained by cyclic permutation of the $a_i$.
The integration contour for the $\lambda_i$ is the interval $[0, 2 \pi]$,
shifted by $- i s_i \beta \zeta$ for $\zeta > 0$ arbitrary
and subject to a pole selection rule which non-trivially deforms the contour.

This deformation of the contour makes it difficult to show that the superconformal index reduces to the
correct expression $Z_{n,0}(q_1, q_2)$ by directly applying the twisted limit.
\paragraph{}
Instead, we perturbatively expand $Z_{S^5 \times S^1}^{U(n)}$ in $e^{-\beta k}$ where
$k$ is the total instanton number,
\begin{equation}
  Z_{S^5 \times S^1}^{U(n)}(\beta, m, a_i) = \sum_{k=0}^{\infty} e^{-\beta k} I_k^{(n)}.
\end{equation}
The twisted limit in terms of the variables $\beta, m, a_i$ is the limit
\begin{equation}
  \begin{aligned}
    \beta &\to \infty, \\
    m &\to 3/2 &&\text{such that} &\tilde{\Delta}_2 &= \beta \left(\frac{3}{2} - m\right) = \text{const}, \\
    a_1 &\to -1 &&\text{such that} &\tilde{\omega}_1 &= \beta (1 + a_1) = \text{const}.
  \end{aligned}
\end{equation}
Additionally, the constraint $a_1 + a_2 + a_3 = 0$ for $a_{2,3} \in (-1,1)$ has to stay satisfied.
The non-vanishing fugacities are
\begin{equation}
  q_2 = e^{- \Delta_2} = e^{- \tilde{\omega}_1} = e^{-\beta (1 + a_1)}
  \quad \text{and} \quad
  q_1 = e^{- \Delta_1} = e^{\tilde{\Delta}_2} = e^{- \beta (m - 3/2)}.
\end{equation}
It might seem that in the twisted limit, since $\beta \to \infty$, all terms with $k > 1$ get infinitely suppressed.
However, the $I_k^{(n)}$ diverge in the same limit in such a way as to render the expansion a finite
expansion in $q_2$. The coefficients are functions of $q = q_1 q_2$, satisfying $|q_2| < |q|$, or $|q_1| > 1$.
This is the manifestation of analytic continuation and seen best when interpreting the expansion in $q_2$ as in $q_1^{-1}$.
\paragraph{Giant Graviton Coefficients}
For ease of comparison, we reproduce the expansion of the $Z_{n,0}(q_1, q_2)$, expanded in variables $q_2$ and $q = q_1 q_2$ with $|q_2| < |q|$, perturbatively in $q_2$.
The result is
\begin{equation}
	\begin{aligned}
		Z_{1,0}(q_1, q_2)      & = 1 + \frac{q_2}{q} + \frac{q_2^2}{q^2} (1+q) +
		\frac{q_2^3}{q^3} (1+q+q^2) + \frac{q_2^4}{q^4} (1+q+2q^2+q^3) + \mathcal{O}(q_2^5),   \\
		Z_{2,0}(q_1, q_2)      & = 1 + \frac{q_2}{q} + \frac{q_2^2}{q^2} (2+q) +
		\frac{q_2^3}{q^3} (2+2q+q^2) + \frac{q_2^4}{q^4} (3+3q+3q^2+q^3) + \mathcal{O}(q_2^5), \\
		Z_{3,0}(q_1, q_2)      & = 1 + \frac{q_2}{q} + \frac{q_2^2}{q^2} (2+q) +
		\frac{q_2^3}{q^3} (3+2q+q^2) + \frac{q_2^4}{q^4} (4+4q+3q^2+q^3) + \mathcal{O}(q_2^5), \\
		Z_{4,0}(q_1, q_2)      & = 1 + \frac{q_2}{q} + \frac{q_2^2}{q^2} (2+q) +
		\frac{q_2^3}{q^3} (3+2q+q^2) + \frac{q_2^4}{q^4} (5+4q+3q^2+q^3) + \mathcal{O}(q_2^5), \\
		                       & \vdots                                                        \\
		Z_{\infty,0}(q_1, q_2) & = 1 + \frac{q_2}{q} + \frac{q_2^2}{q^2} (2+q) +
		\frac{q_2^3}{q^3} (3+2q+q^2) + \frac{q_2^4}{q^4} (5+4q+3q^2+q^3) + \mathcal{O}(q_2^5), \\
	\end{aligned}
\end{equation}
\paragraph{Twisted Limit of Superconformal Index}
The coefficients in the perturbative expansion of $Z_{S^5 \times S^1}^{U(n)}$ up to and including $k = 3$ were calculated in \cite{Kim:2013nva}.
We introduce the variables
\begin{equation*}
	y = e^{\beta(m-1/2)} \quad \text{and} \quad
	y_i = e^{-\beta a_i}.
\end{equation*}
Then,
\begin{equation}
  \begin{aligned}
    I_1^{(n)} &= y && \text{for} \quad n \geq 1, \\
    I_2^{(n)} &= 2 y^2 + y(y_1 + y_2 + y_3) - (y_1^{-1} + y_2^{-1} + y_3^{-1}) + y^{-1}
              &&\text{for} \quad n \geq 2, \\
    I_3^{(n)} &= 
		a_n \, y^3 + 2y^2\sum\limits_{i=1}^3 y_i
		+ y \sum\limits_{i} (y_i^2 - y_i^{-1})
		- \sum\limits_{i \not= j} y_i/y_j
		+ y^{-1} \sum\limits_i y_i
              &&\text{for} \quad n \geq 2,
  \end{aligned}
\end{equation}
where $a_2 = 2$ and $a_3 = 3$. Since the superconformal index coincides with the supergravity index for $k \leq n$,
$a_n = 3$ for $n > 3$.
The abelian $n = 1$ index was matched to all orders in $k$ in \eqref{eq:2510291445}, so we focus on $n \geq 2$.
We immediately find
\begin{equation}
  e^{-\beta} I_1^{(n)} = e^{\beta(m - 3/2)} = q_1^{-1} = \frac{q_2}{q}.
\end{equation}
The terms $e^{- 2 \beta} I_2^{(n)}$ in the twisted limit reduce to
\begin{equation}
  e^{-2 \beta} I_2^{(n)} \longrightarrow 2 e^{2 \beta(m - 3/2)} + e^{\beta(m-3/2)} e^{-\beta (1 + a_1)} =
  \frac{q_2^2}{q^2} (2 + q).
\end{equation}
For the terms of order $e^{-3 \beta}$, we have to distinguish between $n = 2$ and $n \geq 3$. When $n = 2$, the coefficient becomes
\begin{equation}
  e^{-3 \beta} I_3^{(2)} \longrightarrow \frac{q_2^3}{q^3} (2 + 2q + q^2).
\end{equation}
When $n \geq 3$, the difference is an additional $e^{-3 \beta} y^3 = q_2^3 / q^3$, correctly reproducing
\begin{equation}
  e^{-3 \beta} I_3^{(n \geq 3)} \longrightarrow \frac{q_2^3}{q^3} (3 + 2q + q^2).
\end{equation}

\section{Hausel generating function}
\label{sec:fermionic_forms_and_bg_flux}

Consider a quiver with unoriented edges and $K$ nodes.
A $K$-tuple of partitions $\nu = (\nu_0, \ldots, \nu_{K-1}) \in \mathcal{P}^K$ defines the framing of this quiver.
The Hausel function $r$, in the notation of \cite{mozgovoy2007ffqv}, is
\begin{equation}
	\label{eq:hausel_gen_func}
	r(\nu, q^{-1}, \vec{y}) \equiv \sum\limits_{\tau \in \mathcal{P}^K}
  \prod_{k=1}^{\infty} 
  q^{-(\nu_k, \tau_k)}
   q^{\frac{1}{2}(\tau_k, \tau_k)}
   \prod_{\alpha = 0}^{K-1} 
   y_{\alpha}^{\tau_k^{\alpha}}
	\left[\infty, \tau_k^\alpha - \tau_{k+1}^{\alpha} \right]_q
\end{equation}
with the $q$-binomial coefficient
\begin{equation*}
	[n, m]_q \equiv \prod_{i=1}^m \frac{1 - q^{n+i}}{1 - q^i}
	\quad \text{for} \quad
	n \in \mathbb{Z}, m \in \mathbb{N}.
\end{equation*}
We formally set
\begin{equation*}
	[\infty, m]_q = \prod_{i=1}^{m} \frac{1}{1 - q^i}.
\end{equation*}
The partition-vector $\nu$ is interpreted as a list of weight vectors $\nu_k \in \mathbb{N}_0^K$ of the Kac-Moody algebra associated to the quiver diagram (set $\nu_{\alpha,k} = 0$ if the partition $\nu_{\alpha}$ has less than $k$ parts).
Similarly, $\tau$ is interpreted as a list of root vectors $\tau_k$ of the same algebra.
The inner product between weights and roots is just as in (\ref{eq:2511032230}--\ref{eq:2511032231}).
In fact, the only difference between the Hausel function $r$ and the second fermionic form $n$ in (\ref{eq:fermionic_form}) is the appearance of $\infty$ in the $q$-binomial coefficient.
When $\nu$ becomes a $K$-tuple of integers, $\nu \in \mathbb{N}_0^K$, the quiver data with gauge ranks $n^\alpha$, together with the framing $\nu$, defines a quiver variety $\mathcal{M}_{n, \nu}$.
It was shown by Hausel \cite{Hausel_2006} that the ratio
\begin{equation}
  \frac{r(\nu, q^{-1}, \vec{y})}{r(0, q^{-1}, \vec{y})}
\end{equation}
is the generating function of Poincar\'e polynomials $P_q[\mathcal{M}_{n,\nu}]$ of $\mathcal{M}_{n,\nu}$,
\begin{equation}
	\frac{r(\nu, q^{-1}, \vec{y})}{r(0, q^{-1}, \vec{y})} =
  \sum_{n^{\alpha} = 0}^{\infty} 
  P_q [\mathcal{M}_{n, \nu}]
  q^{-d(n, \nu)} \prod_{\alpha = 0}^{K-1} y_{\alpha}^{n^{\alpha}},
\end{equation}
where
\begin{equation}
  d(n, \nu) = n^{\alpha} \nu_{\alpha} - \frac{1}{2} C_{\alpha \beta}
  n^{\alpha} n^{\beta}.
\end{equation}

\paragraph{}
The goal of this section is to show that this ``Hausel generating function'' arises naturally from the monopole formula for the Hilbert series at large rank.
Concretely, we show that the functions $r_l(m, q^{-1}, y)$ as in \eqref{eq:r_l_adhm} and its generalisation $r_{l^0, \ldots, l^{K-1}}(\mathbf{m}, q^{-1}, y)$ as in \eqref{eq:r_l_general} are parts of the Hausel function and that
\begin{equation}
	\label{eq:hausel_gen_func_series}
	\sum_{l^{\alpha}=0}^{\infty}
	r_{l^0, \ldots, l^{K-1}}(\mathbf{m}, q^{-1}, \vec{y}) = r(\mathbf{m}^T, q^{-1}, \vec{y}).
\end{equation}
For convenience, we reproduce the formula for $r_{l^0, \ldots, l^{K-1}}$,
\begin{equation}
	r_{l^0,\ldots,l^{K-1}}(\mathbf{m}, q^{-1}, y) =
  \sum_{\substack{\pi^{\alpha} \in \mathcal{P}\\ l(\pi^{\alpha}) = l^{\alpha}}}
	q^{h(\pi, \mathbf{m})} \prod_{\alpha=0}^{K-1} y_{\alpha}^{|\pi^{\alpha}|} P_{\pi^{\alpha}}(q)
\end{equation}
with
\begin{equation}
	h(\lambda, \mu) = \frac{1}{2} C_{\alpha \beta}
  \sum_{a=1}^{l(\lambda^{\alpha})} \sum_{b=1}^{l(\lambda^{\beta})} B(\lambda_a^{\alpha}, \lambda_b^{\beta})
	- \sum_{\alpha = 0}^{K-1} \sum_{a=1}^{l(\lambda^\alpha)} \sum_{b=1}^{l(\mu_{\alpha})} B(\lambda_a^{\alpha}, \mu_{\alpha,b}).
\end{equation}
The summation over all $l^{\alpha}$ has the effect of removing the restrictions on the partition lengths.
In the resulting summation over all partitions, we transpose them by defining $\tau^\alpha = (\pi^{\alpha})^T$.
This simplifies the expression for $h(\lambda, \mu)$ dramatically. First,
\begin{equation}
	\sum_{a=1}^{l(\lambda)} \sum_{b=1}^{l(\mu)} B(\lambda_a, \mu_b) =
	\sum_{a,b = 1}^{\infty} \min (\lambda_a, \mu_b) =
	\sum_{i,j = 1}^{\infty} \min(i, j) \mu_i(\lambda) \mu_j(\mu),
\end{equation}
where $\mu_i(\lambda)$ is the multiplicity of $j \in \mathbb{N}$ in $\lambda$.
We rewrite this as \cite{Barns-Graham:2018qvx}
\begin{equation}
	\sum_{(i,j,k) \in X} \mu_i(\lambda) \mu_j(\pi)
  \quad \text{where} \quad
	X = \{(i,j,k) \in \mathbb{N}^3 : k \leq i, k \leq j\}.
\end{equation}
$\mu_k(\pi^{\alpha}) = \tau_k^{\alpha} - \tau_{k+1}^{\alpha}$, such that, changing the order of summation, this becomes
\begin{equation}
	\sum_{k=1}^{\infty} \sum_{i,j = k}^{\infty} \mu_i(\lambda) \mu_j(\mu) =
	\sum_{k=1}^{\infty} \lambda_k^{T} \mu_k^{T}
\end{equation}
in terms of their transposed partitions, whose parts satisfy
\begin{equation}
	\lambda_k^{T} = \sum_{i = k}^{\infty} \mu_i(\lambda).
\end{equation}
Therefore,
\begin{equation}
	h(\pi, \mathbf{m}) = \frac{1}{2} C_{\alpha \beta} \sum_{k = 1}^{\infty} \tau_k^{\alpha} \tau_k^{\beta}
	- \sum_{k = 1}^{\infty} \tau_k^{\alpha} \mathbf{m}_{\alpha,k}^T
  =
	\sum_{k = 1}^{\infty} \frac{1}{2} (\tau_k, \tau_k) - (\mathbf{m}_k^T, \tau_k).
\end{equation}
What is left to show is that
\begin{equation}
	P_{\pi^{\alpha}}(q) = \prod_{k = 1}^{\infty} [\infty, \tau_k^{\alpha} - \tau_{k+1}^{\alpha}]_q,
\end{equation}
but this follows straightforwardly since $\mu_k(\pi^{\alpha}) = \tau_k^{\alpha} - \tau_{k+1}^{\alpha}$.
Putting the individual parts together concludes the proof of eq.~\eqref{eq:hausel_gen_func_series}.

\section{Fermionic forms for $K = 1$}
\label{sec:fermionic_forms-adhm_quiver}

In this section, we perform an explicit resummation of the second fermionic form in the plethystic case $K = 1$ (and arbitrary $L$).
The mirror quiver, which underlies the fermionic form, has just a single gauge node with an edge connecting to itself, so the adjacency matrix is $C = (0)$.
Then, the fermionic form simplifies to
\begin{equation}
  \label{eq:2511031456}
	n(\nu, q, y) = \sum_{\tau \in \mathcal{P}} y^{|\tau|} \prod_{k = 1}^{\infty} q^{- \nu_k \tau_k} \left[\sum_{l=1}^k \nu_l, \tau_k - \tau_{k+1} \right]_q.
\end{equation}
Both $\nu$ and $\tau$ are partitions.
Denote the partial sums of $\nu$ by
\begin{equation}
  \tilde{\nu}_k \equiv \sum_{l=1}^k \nu_l.
\end{equation}
We will now show that
\begin{equation}
	\label{eq:poincare_adhm}
	n(\nu, q, y) = \prod_{k = 1}^{\infty}\prod_{l=0}^{\tilde{\nu}_k} \frac{1}{1 - y^k q^{-l}}.
\end{equation}
For the proof of this, we will first transpose all partitions $\tau$ in the summation.
Note that the transpose partition $\tau^T$ to a partition $\tau$ satisfies
\begin{equation}
	\tau_k^{T} = \sum_{l = k}^{\infty} \mu_l(\tau),
\end{equation}
where $\mu_l(\tau)$ is the multiplicity of $l \in \mathbb{N}$ in $\tau$.
Then, transposing the partitions $\tau$ in the summation,
\begin{equation}
	\begin{aligned}
		n(\nu, q, y) & = \sum_{\tau \in \mathcal{P}} y^{|\tau|}
		\prod_{k = 1}^{\infty} q^{-\nu_k \sum_{l = k}^{\infty} \mu_l(\tau)}
		[\tilde{\nu}_k, \mu_k(\tau)]_q                          \\
		             & =
		\sum_{\tau \in \mathcal{P}} y^{|\tau|}
		\prod_{k = 1}^{\infty} [\tilde{\nu}_k, \mu_k(\tau)]_{q^{-1}}  \\
		             & =
		\prod_{k = 1}^{\infty}
		\left[
		\sum_{\mu_k = 0}^{\infty} y^{k \mu_k}
		{\tilde{\nu}_k + \mu_k \choose \mu_k}_{q^{-1}}  \right] \\
		             & =
		\prod_{k = 1}^{\infty}
		\prod_{l=0}^{\tilde{\nu}_k} \frac{1}{1 - y^k q^{-l}}.
	\end{aligned}
\end{equation}
In the second line we first wrote
\begin{equation}
  \prod_{k = 1}^{\infty} q^{-\nu_k \sum_{l=k}^{\infty} \mu_l(\tau)} =
  \prod_{l \geq k}^{\infty} q^{-\nu_k \mu_l(\tau)} =
  \prod_{k = 1}^{\infty} q^{- \tilde{\nu}_k \mu_k(\tau)}
\end{equation}
and then used the identity 
\begin{equation}
  q^{-mn} [m, n]_q = q^{-mn} \prod_{i=1}^{m} \frac{1 - q^{n+i}}{1 - q^i} = [m, n]_{q^{-1}}.
\end{equation}

\paragraph{}
We will now focus on the expressions that come up as the giant graviton coefficients.
First, at $\nu = 0$, the fermionic form simplifies to
\begin{equation}
  n(0, q, y) = \prod_{k=1}^{\infty} \frac{1}{1 - y^k} = \frac{1}{(y; y)_{\infty}}.
\end{equation}
This is as expected from eq.~(\ref{eq:2511031805}) since $K=1$ and the Weyl denominator is trivial.
The ratio of fermionic forms with and without flux is
\begin{equation}
  \label{eq:2511032139}
  \frac{n(\nu, q, y)}{n(0, q, y)} = \prod_{k=1}^{\infty} \prod_{l=1}^{\tilde{\nu}_k} 
  \frac{1}{1 - y^k q^{-l}}.
\end{equation}
The giant graviton coefficients are obtained by setting $\nu = \mathbf{m}^T$, $q = q_1 q_2$ and $y = q_1^L$ or $y = q_2^L$ in (\ref{eq:2511032139}),
\begin{align}
    Z_{\mathbf{0}, \mathbf{m}}(q_1, q_2) &= (q_1^L, q_1^L)_{\infty} \, n(\mathbf{m}^T, q, q_1^L)
    \quad \text{for} \quad \mathbf{m} \in [\mathbf{M}]_+ \quad \text{and} \\
    Z_{\mathbf{m}, \mathbf{0}}(q_1, q_2) &= (q_2^L, q_2^L)_{\infty} \, n(\mathbf{m}^T, q, q_2^L)
    \quad \text{for} \quad \mathbf{m} \in [-\mathbf{M}]_+.
\end{align}
We will explicitly write this out when $\mathbf{M} = \mathbf{0}$.
Then, $\mathbf{m} = (m, \ldots, m) \in \mathbb{N}^L$ and $\mathbf{m}^T = (L, \ldots, L) \in \mathbb{N}^m$.
The coefficients become
\begin{equation}
	\frac{n(\mathbf{m}^T, q, y)}{n(0, q, y)} = \prod_{k=1}^{m} \prod_{l=1}^{L k} \frac{1}{1 - y^k q^{-l}}
	\times
	\prod_{k=m+1}^{\infty} \prod_{l=1}^{L m} \frac{1}{1 - y^k q^{-l}}.
\end{equation}
Specialising on branes wrapped around $z_1 = 0$, the fugacities are $q = q_1 q_2$ and $y = q_2^L$.
The product over $l$ is split up into $l = L k_1 + k$ with $k_1 = 1, \ldots, k_2$ and $k = 0, \ldots, L-1$.
Concretely, the two factors become
\begin{equation}
  \prod_{k_2=1}^{m} \prod_{k_1 = -k_2}^{-1} \prod_{k = 0}^{L-1} \frac{1}{1 - q_2^{L k_2 + L k_1 + k} q_1^{L k_1 + k}}
\end{equation}
and
\begin{equation}
  \prod_{k_2=m+1}^{\infty} \prod_{k_1 = -m}^{-1} \prod_{k = 0}^{L-1} \frac{1}{1 - q_2^{L k_2 + L k_1 + k} q_1^{L k_1 + k}}.
\end{equation}
Exchanging the order of $k_1$ and $k_2$ in the products, they combine into
\begin{equation}
  \prod_{k=0}^{L-1} \prod_{k_1 = -m}^{-1} \prod_{k_2 = -k_1}^{\infty} \frac{1}{1 - q_2^{L k_2 + L k_1 + k} q_1^{L k_1 + k}},
\end{equation}
which is simply
\begin{equation}
  \prod_{k=0}^{L-1} \prod_{k_1 = -m}^{-1} \prod_{k_2 = 0}^{\infty} \frac{1}{1 - q_2^{L k_2 + k} q_1^{L k_1 + k}} = Z_{0, m, 0}(q_1, q_2),
\end{equation}
in agreement with eq.~(\ref{eq:2511032151}).


\bibliographystyle{JHEP}
\bibliography{biblio.bib}

\end{document}